\documentclass[twocolumn]{aastex62}
\usepackage{savesym}
\savesymbol{tablenum}
\usepackage{siunitx}
\restoresymbol{SIX}{tablenum}
\usepackage{amsmath}
\usepackage{mathrsfs}
\usepackage{multirow}
\usepackage[english]{babel}
\usepackage{blindtext}

\DeclareSIUnit\rg{r_g}
\DeclareSIUnit\kev{\kilo\electronvolt}

\newcommand{\rbp}{r$_{\rm BP}$}
\newcommand{\risco}{r$_{\rm ISCO}$}
\newcommand{\iin}{$i_{\rm in}$}
\newcommand{\iout}{$i_{\rm out}$}
\newcommand{\eq}[2]{\mbox{#1 $=$ #2}}

\newcommand{\xillver}{\textsc{xillver}}
\newcommand{\xillverrr}{\textsc{xillver}\textrm{RR}}
\newcommand{\xspec}{\textsc{xspec}}
\newcommand{\relxill}{\texttt{relxill}}
\newcommand{\relxilllp}{\texttt{relxill\_lp}}

\newcommand{\h}{$h$}
\newcommand{\err}[3]{$#1_{-#2}^{+#3}$}

\graphicspath{{./}{figures/}}

\shorttitle{Warped Disk Iron Line}
\shortauthors{Abarr \& Krawczynski}

\begin{document}

\title{The Iron Line Profile from Warped Black Hole Accretion Disks}
\correspondingauthor{Quincy Abarr}
\email{qabarr@wustl.edu}

\author{Quincy Abarr}
\author{Henric Krawczynski}
\affil{Washington University in St. Louis and McDonnell Center for the Space Sciences}

\begin{abstract}
The profile of the fluorescent iron line from black hole accretion disks is a powerful diagnostic of black hole properties, such as spin and inclination.
The state-of-the-art, however, considers an accretion disk whose angular momentum is aligned with that of the black hole; this is a very constraining assumption which is unlikely to apply to many astrophysical systems.
Here, we present the first simulation of the reflection spectrum from warped accretion disks using a realistic model of the reflected emission based on the \xillver{} code. 
We present the effects that the radial location of the warp and the tilt angle have on the line profile, showing that the affect becomes significant at relatively low angles, between \SIlist{5;15}{\degree}.
We highlight that the results are highly dependent on the azimuth position of the observer relative to the tilt angle.
We fit these profiles in \xspec{} with the standard \relxill{} lamppost model to quantify the effect that neglecting the disk warps has on the inferred black hole spins and inclinations, finding that the spin parameter can be off by as much as \num{0.2}. We show that fits with two-component \relxill{} can be used to derive more accurate parameter estimates and can recover the radial location of the warp.
\end{abstract}


\section{Introduction} \label{sec:intro}

Emission lines are a powerful tool for probing the geometry and structure of the inner accretion flows of compact objects. 
In the case of stellar mass and supermassive black holes, the fluorescent iron K$\alpha$ emission has proven to be a particular powerful diagnostic \citep[e.g.][]{rn03, brenneman2006, mcclintock2011}. 
In some objects, this line is broadened by the gravitational redshift from the potential well and Doppler shift from the orbital motion of the disk plasma. 
The line shape depends on many parameters, including the emissivity profile of the disk, the ionization state of the disk material, the inner edge of the disk, and the inclination of the observer relative to the angular momentum vector of the disk.
When studying the spectra of black holes (BH), the inclination and spin of the BH are often inferred from a fit to the data assuming that the disk is geometrically thin with parallel BH spin and accretion disk spin axes. 
This assumption, although common, may not be accurate in a significant fraction of observed objects. 
A misalignment of the BH spin and the binary orbit (stellar mass black holes) or the inner gas flow (supermassive black holes) can lead to a warped or fragmented accretion disk. 
For example, \cite{bp75} conjectured that the combination of the viscosity of the accreting plasma and the general relativistic Lense-Thirring precession will align the inner accretion disk axis with the BH spin axis.
In x-ray binaries (XRBs), the supernova may kick the BH out of alignment with the companion star as it forms; \cite{brandt1995} estimate that as many as \SI{60}{\percent} may have misalignments between \SIrange[range-phrase={ and }]{5}{45}{\degree}.
\cite{fragos2010} make more pessimistic predictions using population synthesis models, estimating that more than two-thirds of BHs in XRBs have misalignments smaller than \ang{10}, and that only about \SI{5}{\percent} have misalignments greater than \ang{20}.  
In active galactic nuclei (AGN), the scale of the supermassive black hole (SMBH) is tiny compare to the matter falling onto it; over long timescales the axes of the disk and black hole angular momentum will align, but early on in young or newly-formed galaxies this may not be true.
Galaxies resulting from recent mergers, for example, may contain such misalignments \citep{volonteri2005}. 

Recent general relativistic magnetohydrodynamic (GRMHD) simulations confirm that oblique accretion can indeed lead to a warped accretion disk \citep{liska2018}.
The simulations indicate that the warp can occur closer to the BH than previously thought. 
\cite{liska2019}, for example, showed that around a BH with spin $a=0.9375$, a disk with aspect ratio $H/R=0.03$ and initial misalignment of \ang{10} enters a persistent warped configuration where the disk aligned with the BH inside \SI{~5}{\rg}

It is important to understand how the disk warp impacts the observed properties of the X-ray emission.
\cite{fragile2005} showed that the iron line profiles are sensitive to the tilt and radius of the warp and argued that the broadened iron line of the AGN MCG--6-30-15 may require a warped disk model \citep[see also][]{wang2012}. 
\cite{cheng2016} and \cite{abarr2020} studied the impact of the disk warp on the polarization of the thermal and reflected disk emission. The predicted signatures may become measurable with the Imaging X-ray Polarization Explorer (IXPE) to be launched in 2021 or 2022 \citep{ixpe}. 

One of the strongest cases for the presence of warped disk is through their prediction as a source of quasi-periodic oscillations (QPOs).
\cite{miller2005} first found that the strength of the flux and width of the Fe K$\alpha$ line changes with the phase of the \SIlist{1;2}{\hertz} QPO in the microquasar GRS 1915+105.
\cite{schnittman2006} fit this data with a precessing ring model located at \SI{\sim10}{\rg} and an inclination of \ang{\sim70}.
This ring model is distinct from the Bardeen-Petterson effect; it is instead a precessing thick inner region, with the rest of the disk aligned out to \SI{100}{\rg}.
It does, however, support the link between QPOs and an inclined disk region close in to the BH.
\cite{ingram2012} further linked the change in median iron line energy to the frequency of QPOs if the QPO comes from precession of the inclined hot inner flow.

In this paper, we present results from modeling the reflected emission in greater detail than before.
We assume a lamppost geometry to infer the irradiation of the accretion disk with hard X-rays (rather than assuming a power law emissivity profile).
Furthermore, we replace the delta function line emissivity in the plasma frame ($\epsilon\propto \delta(E-6.4\,\rm keV)$) used by \cite{fragile2005} and \cite{wang2012} with the inclination dependent reflection energy spectrum inferred from detailed radiation transport calculations \citep{garcia2013}.
Invoking simplifying assumptions (i.e. neglecting radiative heating of the emitting plasma), we explore the impact of photons that scatter in the warped disk configuration.
For the first time, we present results from fitting the results with the commonly used \relxilllp{} model \citep{garcia2014} and show how the warp impacts the inferred BH disk properties, such as inclination and spin.

The outline of this paper is as follows.
In Section \ref{sec:methods}, we describe the ray-tracing code and define the geometry of the warped accretion disk.
In Section \ref{sec:iron_emission}, we present the simulated energy spectra, including the contribution of photons which scatter multiple times off the disk, and show how the line profiles change as a function of the warp geometry.
In Section \ref{sec:fitting} we fit the simulated energy spectra with \relxilllp{}, emphasizing that some of the energy spectra can be fit rather well with a two-component disk model.
Finally, in Section \ref{sec:conclusion}, we summarize our results.

\section{Methods} \label{sec:methods}

\subsection{General Relativistic Ray-Tracing Code} \label{subsec:code}
Our ray-tracing code generates power law photon beams in a lamppost corona at height \h{} above the black hole and \ang{1} off the spin axis to avoid the coordinate singularity here; the code is also capable of producing thermal photons from the accretion disk, but we do not consider them in this work.
Power law photons are emitted isotropically from the corona in the plasma rest frame.
The wave vector of each photon is transformed into the global Boyer-Lindquist coordinate system, after which the geodesic is integrated using the Cash-Karp method \citep{ck1990} until the photon comes within 2\% of the event horizon of the BH or reaches a coordinate stationary observer located at \SI{10000}{\rg}.
In the former case, the photon is considered to be lost within the black hole; in the latter, the wave vector is transformed into the reference frame of the observer for analysis.

When a photon beam crosses the accretion disk it is reflected by transforming its wave vector into the accretion disk frame, using the results of \cite{chandrasekhar} for reflecting photons off a indefinitely thick electron gas, and transforming the beam back into the global Boyer-Lindquist coordinate frame.
One of the advantages of tracing photon beams forward in time is that it allows us to consider these reflections, which is essential to processes such as fluorescence and polarization.

Our code defines the accretion disk as infinitesimally thin, extending from the innermost stable circular orbit, or ISCO (which is dependent upon the spin of the BH) out to \SI{100}{\rg}.
Within the radius \rbp{}, the disk lies in the equatorial plane; outside of this radius, it lies in the plane defined by the solutions to
\begin{equation}
  \label{eq:disk_plane}
  \cos(\theta)\cos(\beta)=\sin(\theta)\cos(\phi)\sin(\beta),
\end{equation}
where $\beta$ is the misalignment of the outer disk (\eq{$\beta$}{\ang{0}} being no misalignment).

\begin{figure}
    \centering
    \includegraphics[width=\linewidth]{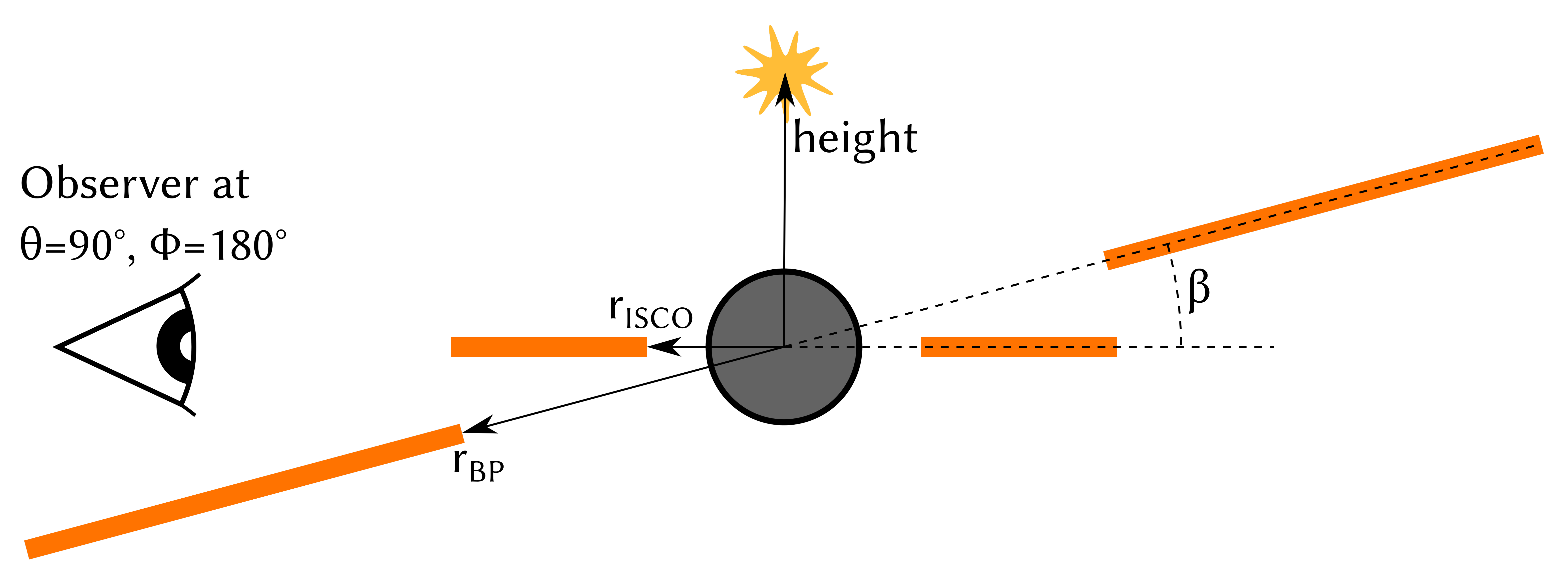}
    \caption{The warped disk model we use is defined by four parameters: the BH spin $a$ (which determines \risco, the radius \rbp{} at which the disk transitions from aligned with the BH spin axis to misaligned, the angle of misalignment $\beta$, and the height \h{} of the lamppost corona. This image of the disk is from an observer located at \eq{$\theta$}{\ang{90}}, \eq{$\phi$}{\ang{270}}.}
    \label{fig:disk}
\end{figure}

This configuration is shown in Figure \ref{fig:disk}.

At all radii, the disk matter is considered to be locally moving in a circular orbit with Keplerian angular velocity $\Omega_K=\left(a+r^{2/3}\right)^{-1}$, following the standard thin disk assumption.
In the Kerr metric, inclined orbits of constant radius precess and have a corkscrew-like shape; we assume, however, that viscous stress forces the outer disk material into circular orbits of constant $r$, and that we can thus use the Keplerian assumption for the outer disk.
The ascending node of the orbiting outer disk material is at \eq{$\phi$}{\ang{90}} and the descending node is at \eq{$\phi$}{\ang{270}}.

In our analysis, we tie the $\theta$-position of the observer to the inclination of the inner disk: \eq{\iin{}}{$\theta$}.
This means that the inclination of the outer disk \iout{} varies with the azimuthal position $\phi$ of the observer according to
\begin{equation}
    \label{eq:iout} 
    i_{\rm out} = \arccos\left(\cos i_{\rm in}\cos\beta - \sin i_{\rm in}\cos\phi\sin\beta\right).
\end{equation}

For an overview of the code as originally written, including its ability to account for polarized scattering, see \cite{krawcz2012}. 
For a more detailed description of our warped disk model, including the equations for the basis used to transform the wave vectors of photon beams into the frame of the outer disk, see \cite{abarr2020}.

In the simulations used in this paper, we track photon beams originating in the lamppost corona.
We vary $\beta$, \rbp{}, and \h{}, but for all simulations we use a BH spin of $a=0.9$ since the affect of spin on the iron line profile is well studied \cite[see e.g.][]{rn03}.
Our default simulation parameters are \eq{\rbp{}}{\SI{15}{\rg}}, \eq{$\beta$}{\ang{15}}, and \eq{\h{}}{\SI{5}{\rg}}; if not otherwise stated, these will be the parameters used.
For each disk configuration, we simulate a total of \num{3.5E8} photons.
Photons which arrive at the observer at \SI{10000}{\rg} are collected in thirty-two bins located at \eq{$\theta$}{\SIlist{25;40;60;75}{\degree}} and \eq{$\phi$}{\SIlist{0;45;90;135;180;225;270;315}{\degree}}. 
These bins are \ang{\pm4} wide in both $\theta$ and $\phi$, and we refer to each of these as individual ``observers''.
The number of photons collected by each observer increases with $\theta$ (and varies less significantly with $\phi$), with the \eq{$\theta$}{\ang{25}} observers seeing \num{\sim1.1E5} photons and the \eq{$\theta$}{\ang{75}} observers seeing \num{\sim3.8E5}.

\subsection{Iron Line Emission} \label{sec:emission}
\begin{figure}
    \centering
    \includegraphics[width=\linewidth]{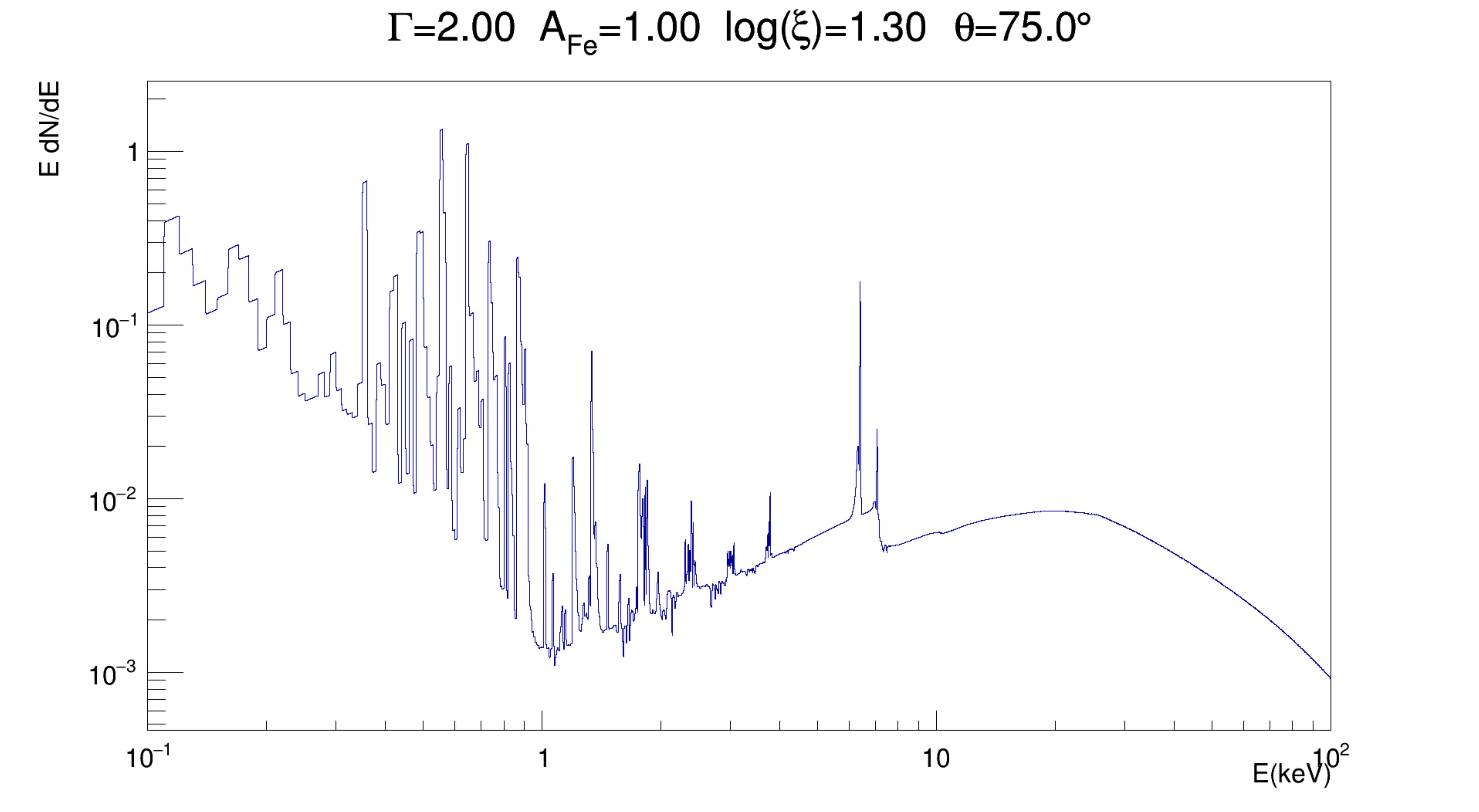}
    \caption{Exemplary \xillver{} energy spectrum used to generate photons in the rest frame of the disk plasma for photons emitted at an inclination of \ang{75}. We use typical disk parameters for an AGN accretion disk: \eq{$\Gamma$}{2}, \eq{$A_{Fe}$}{1}, \eq{$\log(\xi)$}{1.3}.}
    \label{fig:xillver_spectra}
\end{figure}

For the iron emission in the plasma frame of the accretion disk, we use the \xillver{} library, detailed in \cite{garcia2013}.
We have used \xillver{} to create the reflected spectrum centered on the rest-frame iron line energy (\SI{~6.4}{\kev}) for disk inclinations at each integer angle between \SIrange[range-phrase={ and }]{0}{90}{\degree}; the iron line flux changes by about an order of magnitude over this range.
In their Figure 11, \cite{garcia2013} shows Fe K spectra for different ionization parameters; we take the appropriate spectrum from \xillver{} based on the ionization state.
In this paper we focus on the reflected emission from AGN (though this procedure would also be valid for binary BHs by using different parameters), so we use a lamppost power law index $\Gamma=2$ and an ionization parameter $\log(\xi)=1.3$; for a disk viewed at \ang{75} inclination, this spectrum is shown in Figure \ref{fig:xillver_spectra}.

For every photon beam seen by an observer during analysis, we take the \xillver{} spectrum for the emission angle of that beam in the plasma frame of its final reflection, and shift the profile based on its frequency shift over the geodesic between emission and detection at the observer, including transformation out of the plasma frame and into the coordinate stationary observer frame. 

\section{Iron Emission from Warped Disks} \label{sec:iron_emission}

\begin{figure}[t]
    \centering
    \includegraphics[width=0.8\linewidth]{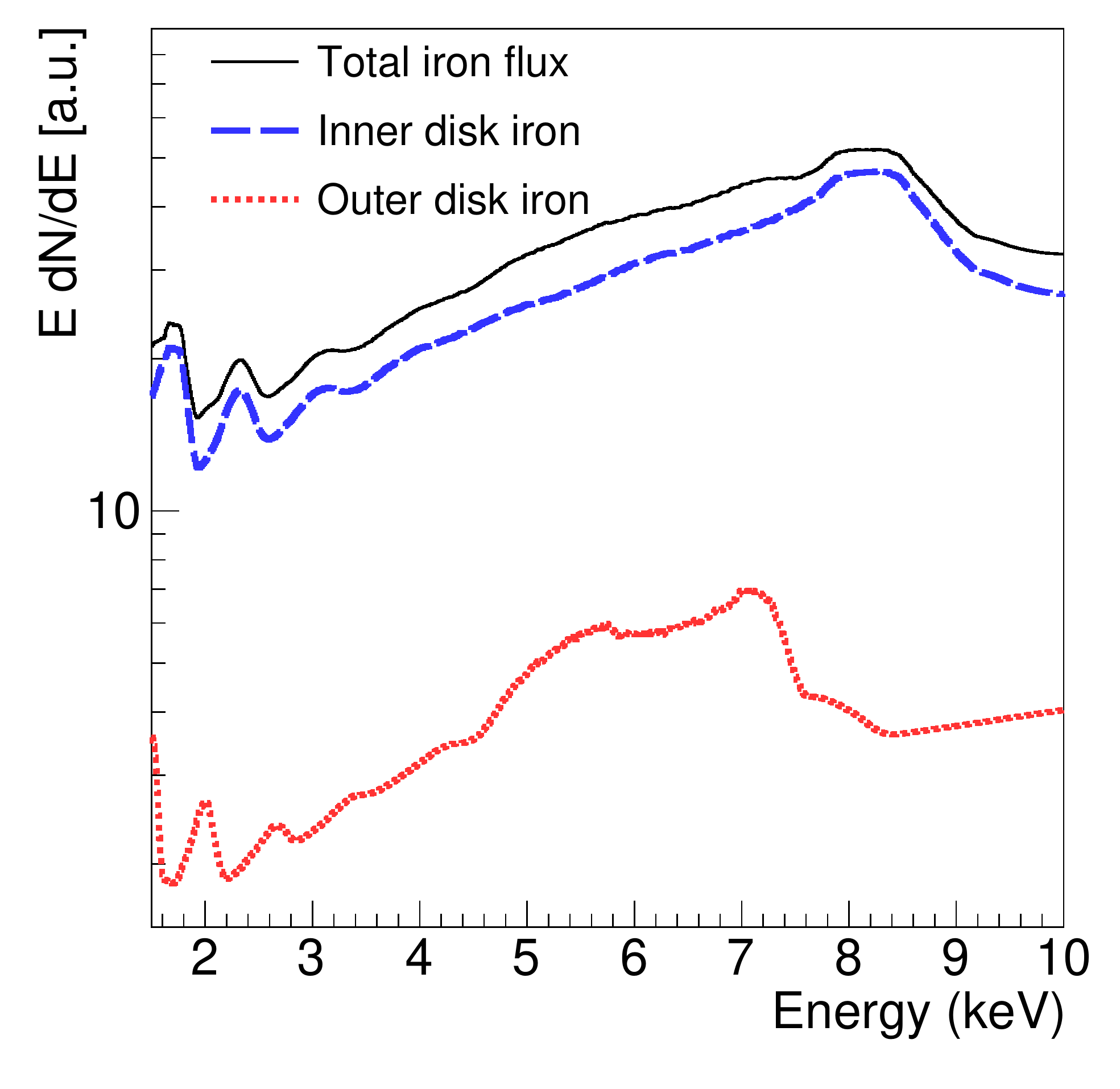}
    \caption{For a simulation with \eq{\rbp{}}{\SI{15}{\rg}}, \eq{$\beta$}{\ang{15}} and an observer located at \eq{$\phi$}{\ang{270}}, \eq{$\theta$}{\ang{75}}, the total reflected flux (in black) is broken down into contributions from the photons reflected off the inner disk (blue) or outer disk (red).}
    \label{fig:profile_components}
\end{figure}

In Figure \ref{fig:profile_components} we break down the iron line profile for our standard simulation observed from \eq{$\phi$}{\ang{270}}, \eq{$\theta$}{\ang{75}} into the contributions from photon beams which scatter once off either the inner or outer disk.
We see that each contributes a blue and red horn, and so even though the blueshifted peak of the inner disk is the dominant feature, there is an enhancement in the profile at the peak red and blue energies from the outer disk.

\subsection{Multiply scattering photons} \label{sec:multiple}
Previous studies have used backwards ray-tracing codes, which cannot account for photon beams which are incident on the disk multiple times before arriving at the observer.
With our forward ray-tracing code, we can track photon beams through an arbitrary number of scattering events before reaching the observer, allowing us to investigate the effect this has on the iron line profile.
The question of how to weight photons which scatter multiple times is a difficult one.
If we consider a beam which scatters twice, for example, at the second scattering the incident flux would not be a pure power law, like \xillver{} assumes; it would instead be a combination of the power law and the fluorescence from the first scattering.

To get a rough approximation of the magnitude of multiple scatterings, we do the following.
For photons which scatter multiple times, we take the \xillver{} spectrum of the first scattering event.
Then, we apply the frequency shift of the geodesic from the final scattering event to the the coordinate stationary observer.
This profile will obviously not be accurate, but should give us a useful estimate of how much of a correction the inclusion multiple scatterings would be.

\begin{figure}[t]
    \centering
    \includegraphics[width=0.8\linewidth]{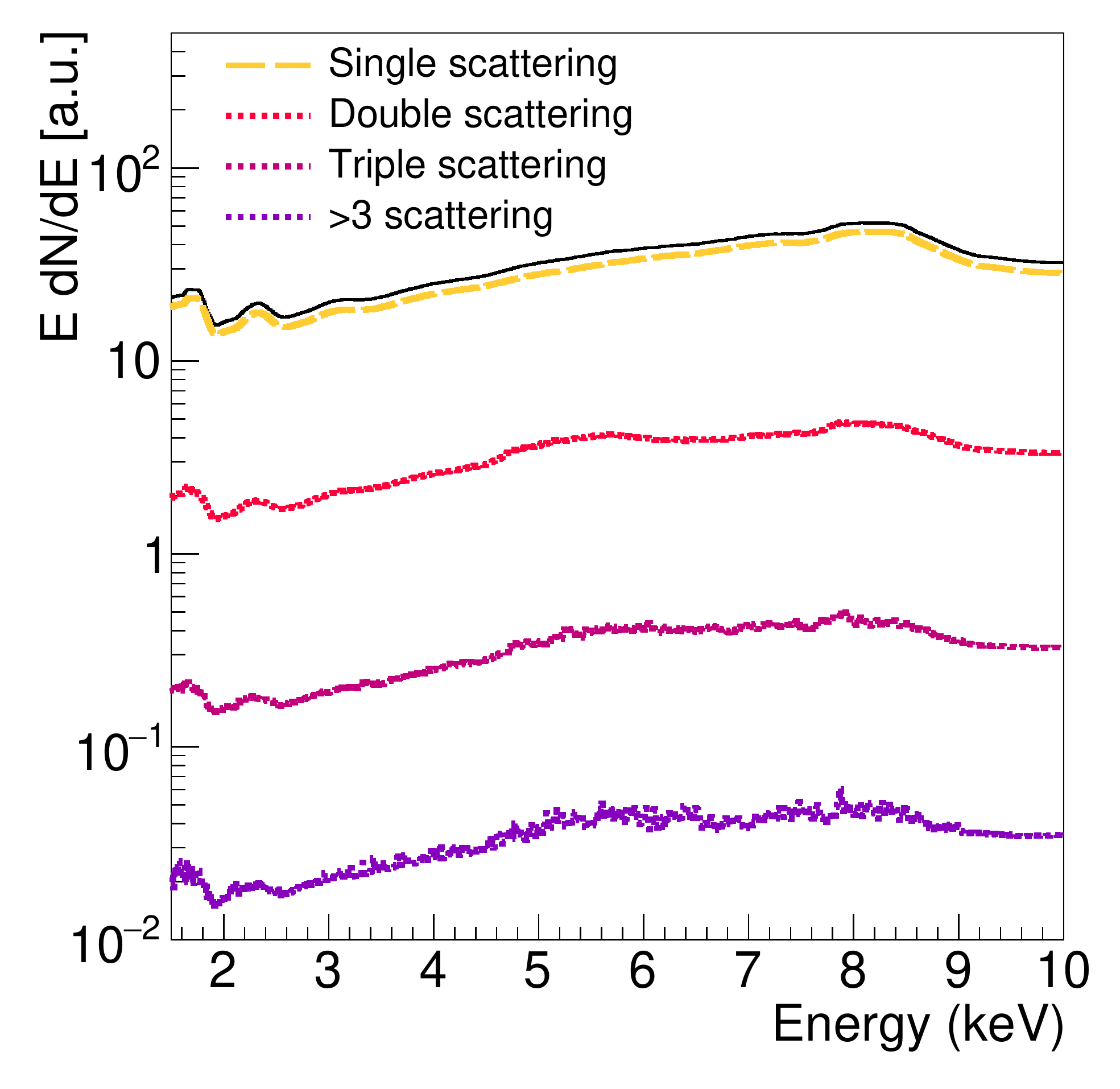}
    \caption{For a simulation with \eq{\rbp{}}{\SI{15}{\rg}}, \eq{$\beta$}{\ang{15}} and an observer located at \eq{$\phi$}{\ang{270}}, \eq{$\theta$}{\ang{75}}, the total reflected flux (in black) is broken down based on the number of times the contributing photons scatter.}
    \label{fig:profile_scatterings}
\end{figure}

\begin{figure}[h]
    \centering
    \includegraphics[width=0.49\linewidth]{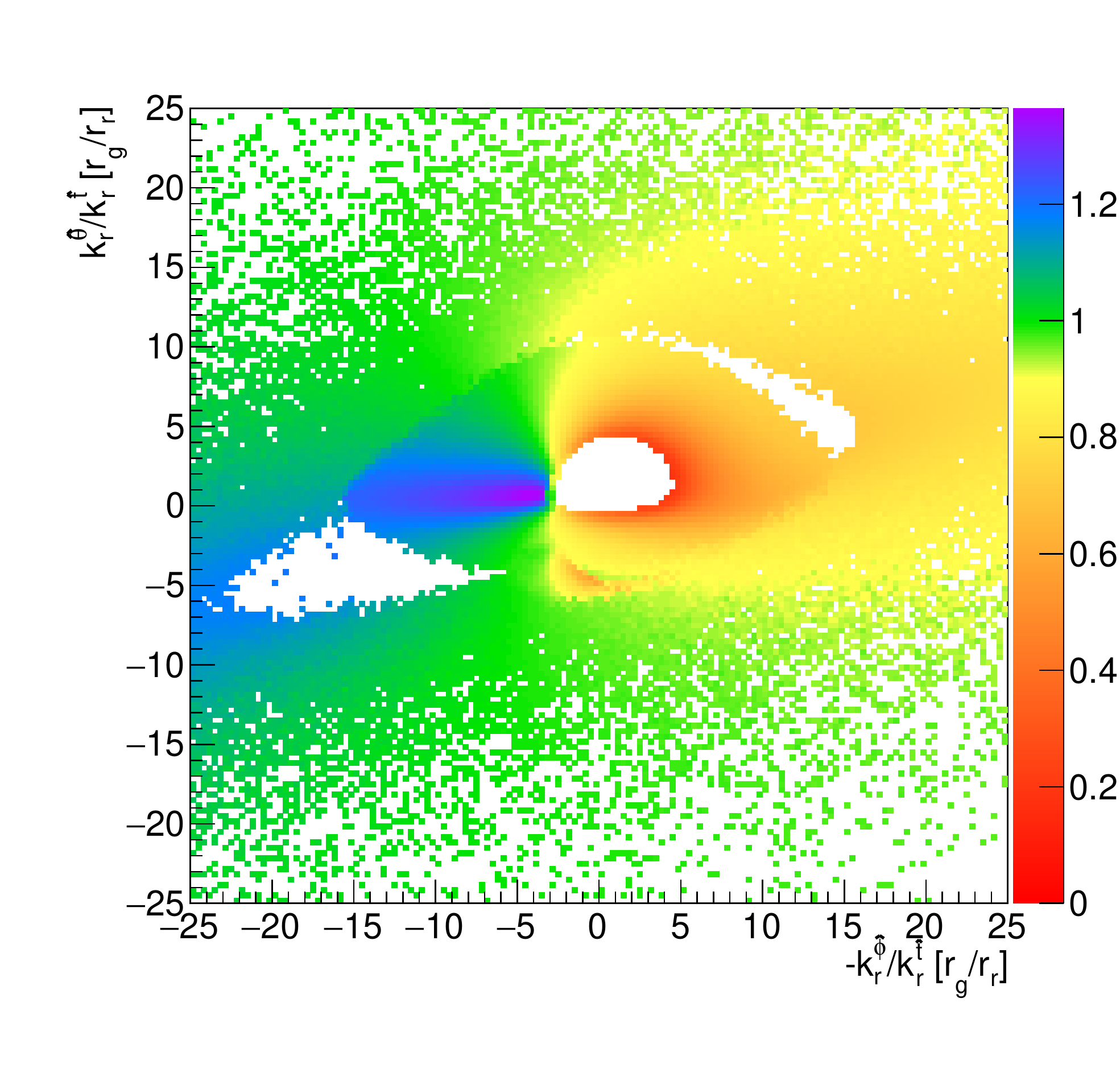}
    \includegraphics[width=0.49\linewidth]{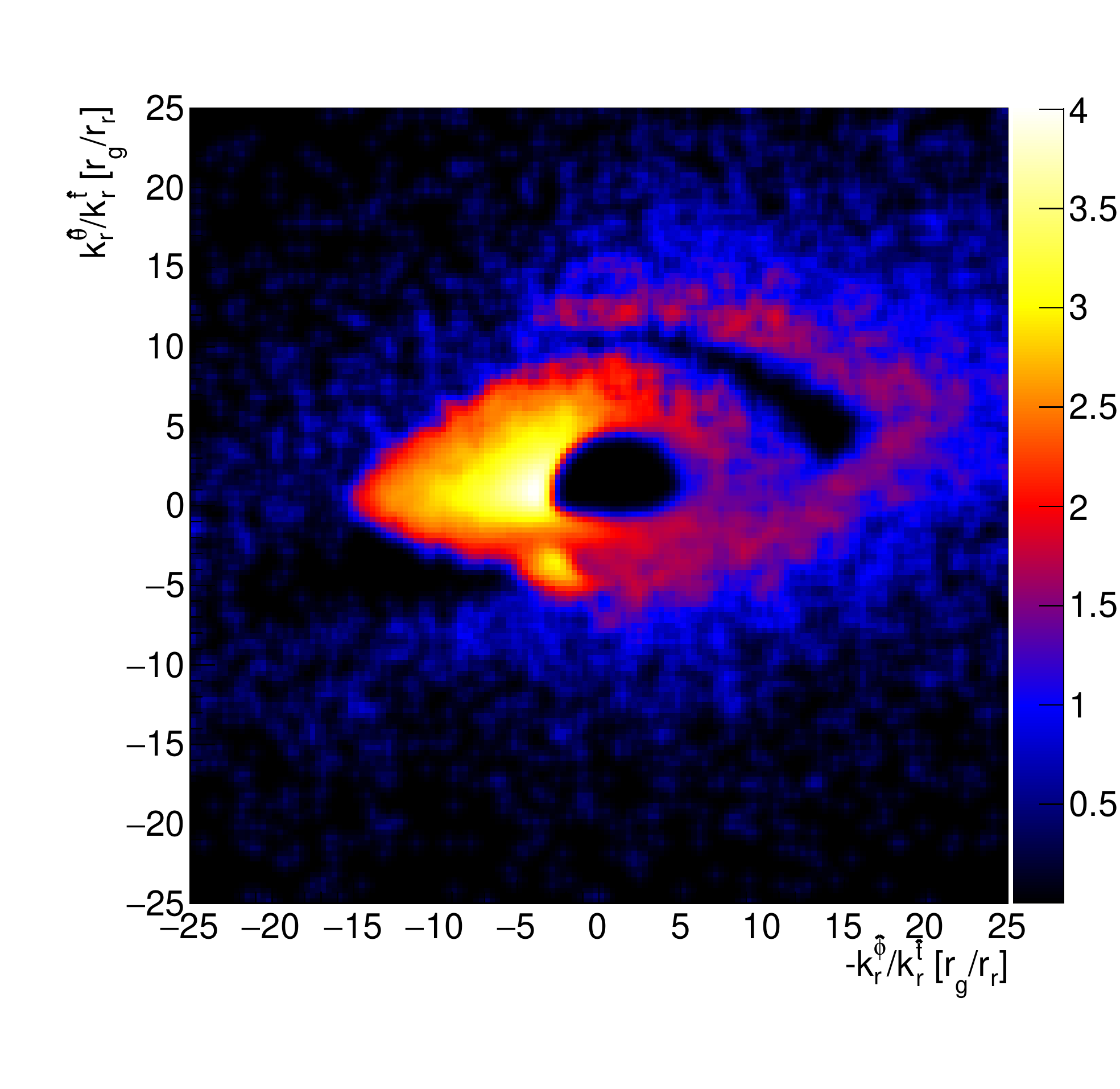}
    \caption{Left: Map of the net frequency shift of photon beams between emission and detection, and Right: intensity map of the reflected emission from two or more scatterings (right, logarithmic scale). Both images are from an observer at \SI{10000}{\rg}, \eq{$\theta$}{\ang{75}} and \eq{$\phi$}{\ang{270}}.}
    \label{fig:map}
\end{figure}

In Figure \ref{fig:profile_scatterings}, we show result from this analysis.
There is an approximately order of magnitude drop in flux with each subsequent scattering event.
The profiles are mostly similar in shape, although the multiple scatterings show a slight enhancement around \SI{5}{\kev}, likely from scattering off the outer disk since this is the peak of the redshifted energy of the outer disk contribution in Figure \ref{fig:profile_components}.
To investigate this, we can look at the maps of the accretion disk.
In Figure \ref{fig:map}, we show two images of the disk: on the left is the net frequency shifts between emission in the plasma frame and detection in the observer frame for all detected photon beams, and on the right is the apparent surface brightness of the disk, but only from photon beams which scatter multiple times.
We see a significant asymmetry in the intensity from the outer disk, with the majority coming from the part of the disk inclined above the equatorial plane.
This is likely because photons do not have to bend as much when they travel over the BH to encounter the inclined outer disk as they would to encounter an unwarped disk in the equatorial plane.
This enhanced region of the outer disk is orbiting away from observer, and based on the left hand plot in Figure \ref{fig:map} this region has a shift of \num{\sim0.8}, which lowers the iron peak to \SI{\sim5}{\kev}.

\begin{figure}[t]
    \centering
    \includegraphics[width=\linewidth]{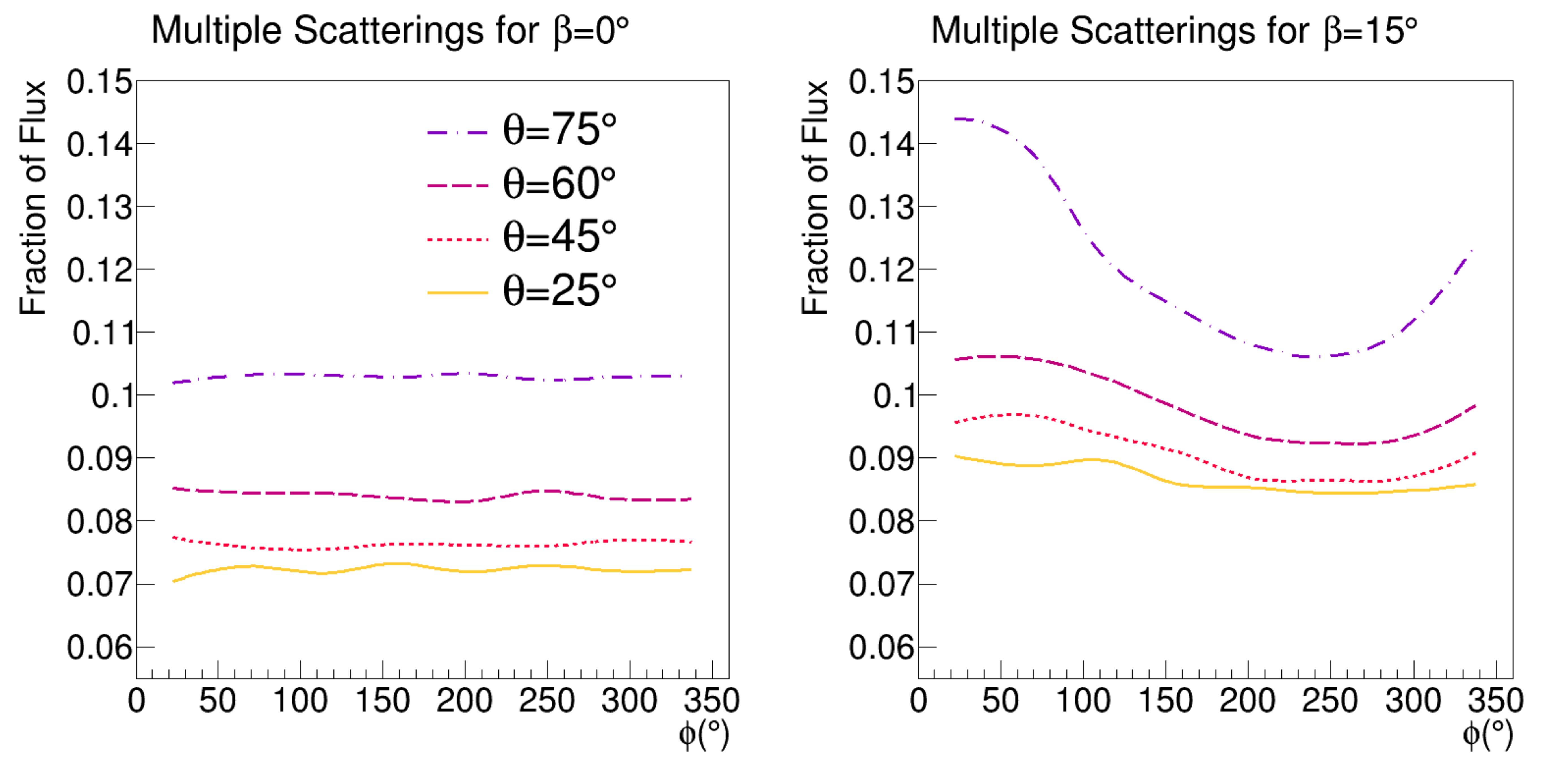}
    \caption{Left: The contributions of multiply scattered photons to the flux for an unwarped disk and Right: the contributions for a warped disk with a \ang{15} tilt, both as a function of the observer's inclination.}
    \label{fig:multi_cont}
\end{figure}

In Figure \ref{fig:multi_cont}, we compare the contributions to flux in the unwarped and warped cases.
On the left we plot the fraction of flux between \SIrange[range-phrase={ and }]{1.5}{10}{\kev} from multiply scattered photons in an unwarped disk.
At the highest inclination shown (\ang{75}), multiple scatterings accounts for about \SI{10}{\percent} of the total flux.
For a warped disk, shown on the right in Figure \ref{fig:multi_cont}, flux fraction modulates with $\phi$ (thus with outer disk inclination), and the degree of modulation increases with the inner disk inclination.
The contribution increases by almost \SI{5}{\percent} at the highest inclination when \eq{$\phi$}{\ang{0}}.

\begin{figure}
    \centering
    \includegraphics[width=0.49\linewidth]{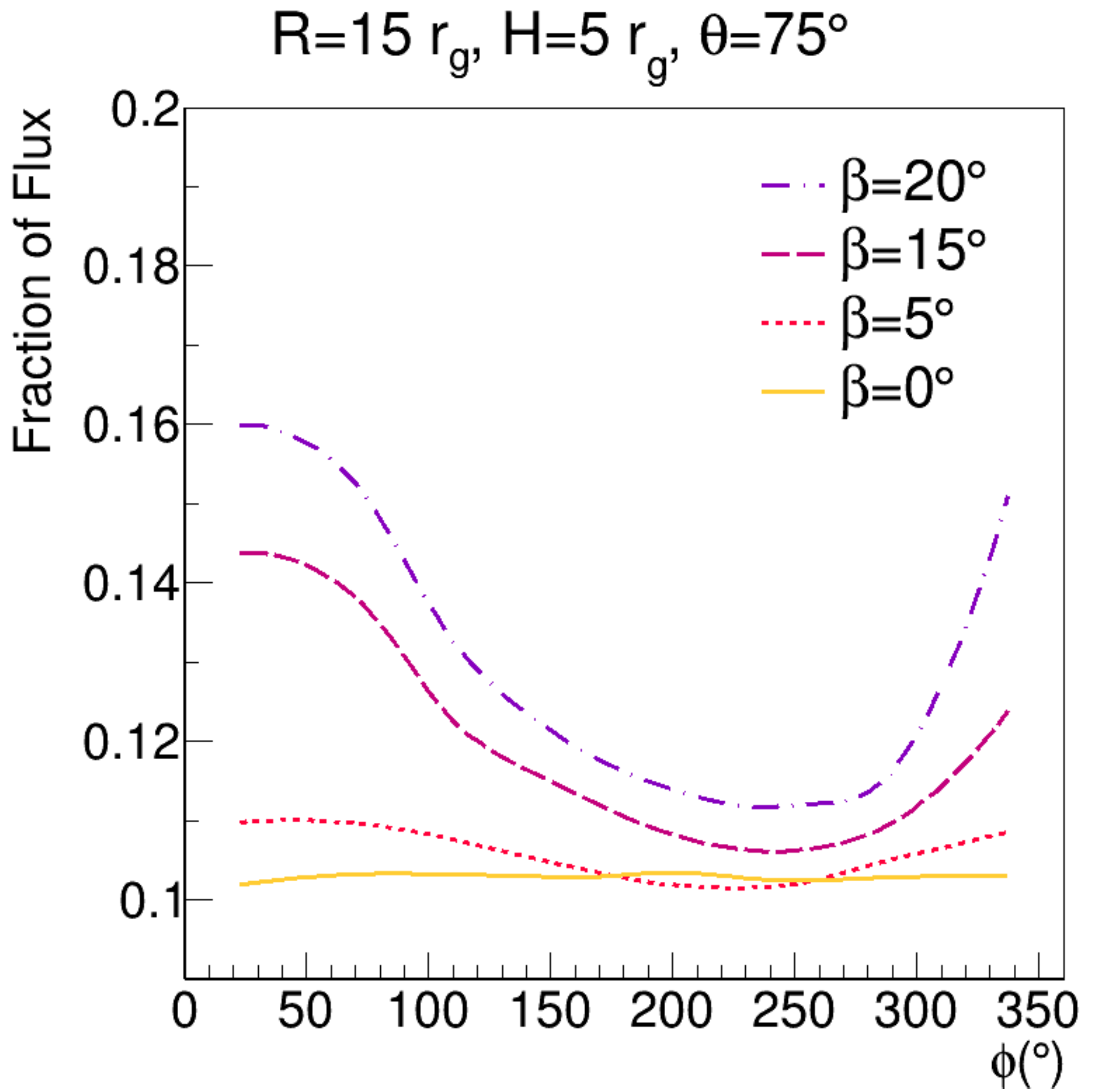}
    \includegraphics[width=0.49\linewidth]{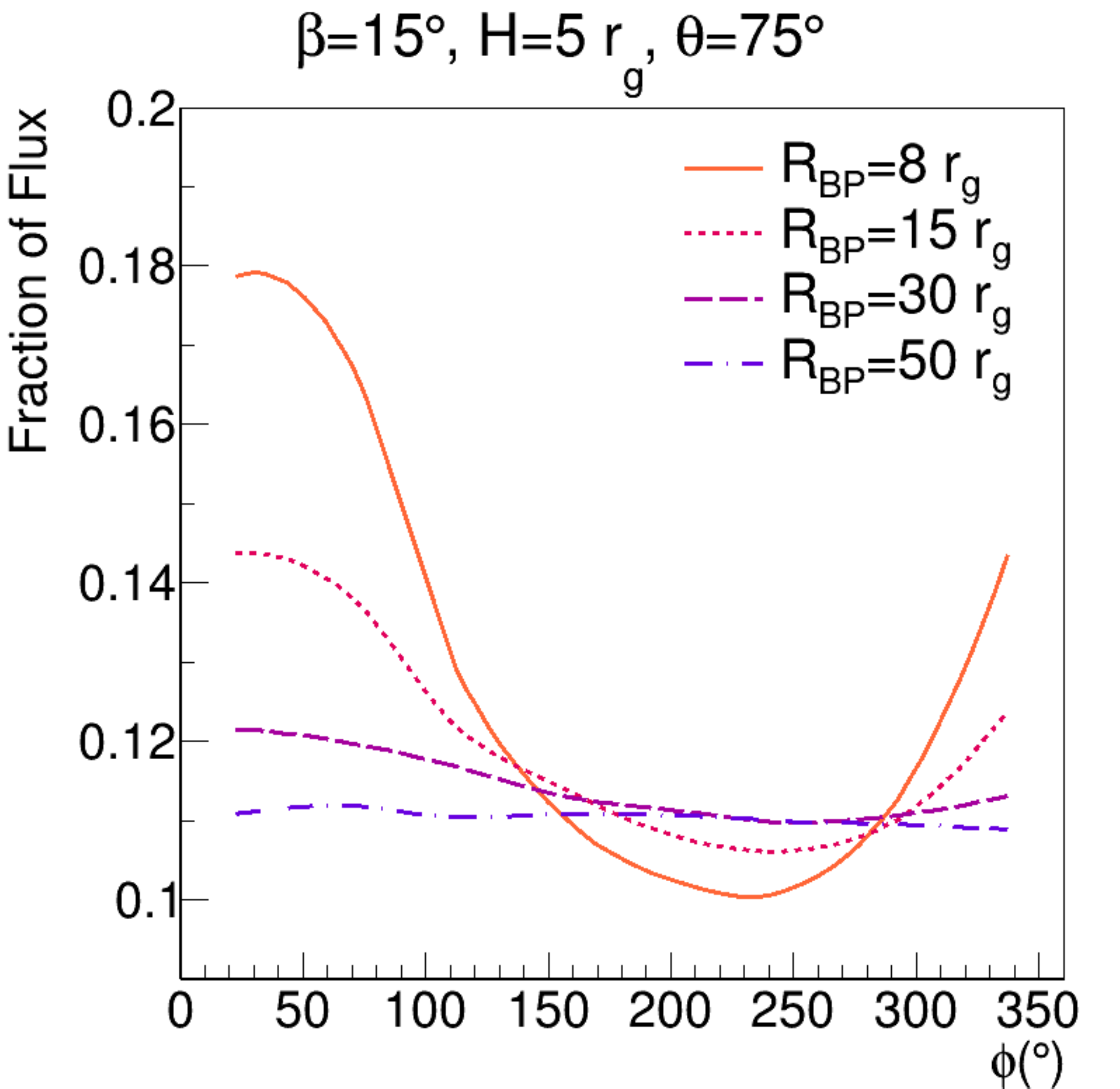}
    \caption{Left: The fraction of flux contributed by photons scattering multiple times for several values of disk warp, and Right: for several values of \rbp{}.}
    \label{fig:multi_comparison}
\end{figure}

To see how the flux fraction varies with the warped disk geometry, we have plotted the fraction at \ang{75} inclination for simulations with varying values of \rbp{} and tilt in Figure \ref{fig:multi_comparison}.
We find that the contribution is proportional to $\beta$ and inversely proportional to \rbp{}, with the contribution increasing to \SI{\sim18}{\percent} for \eq{\rbp{}}{\SI{8}{\rg}}, almost twice the value for an unwarped disk.
Though the specifics of these results are based on a rough estimation, clearly disk warping increases the importance of multiple scattering photons, especially when the disk tilt is large and warp radius is small.

\subsection{Shadowing of the Outer Disk} \label{sec:shadow}

\begin{figure*}
  \centering
  \includegraphics[width=0.4\linewidth]{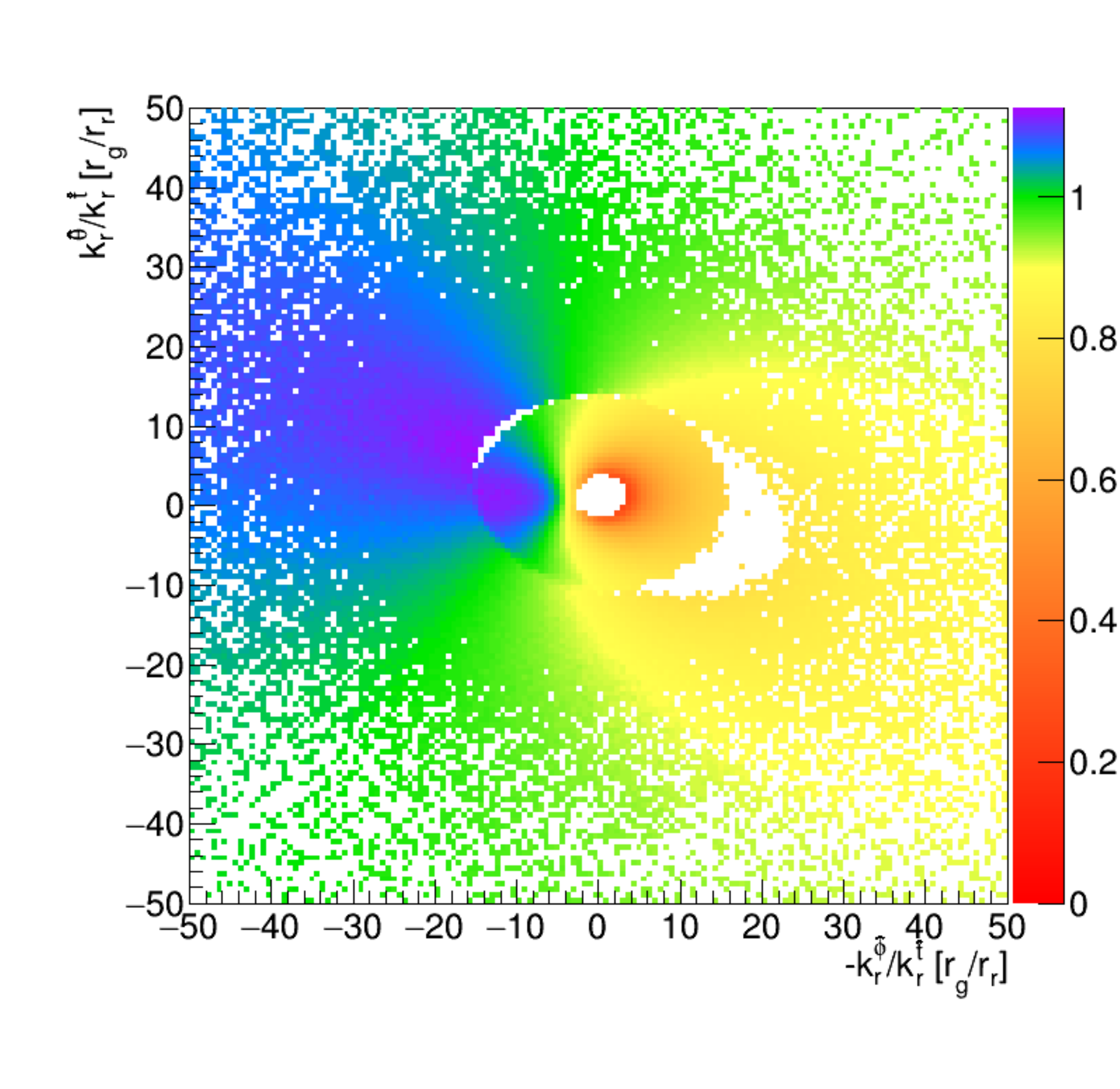}
  \includegraphics[width=0.4\linewidth]{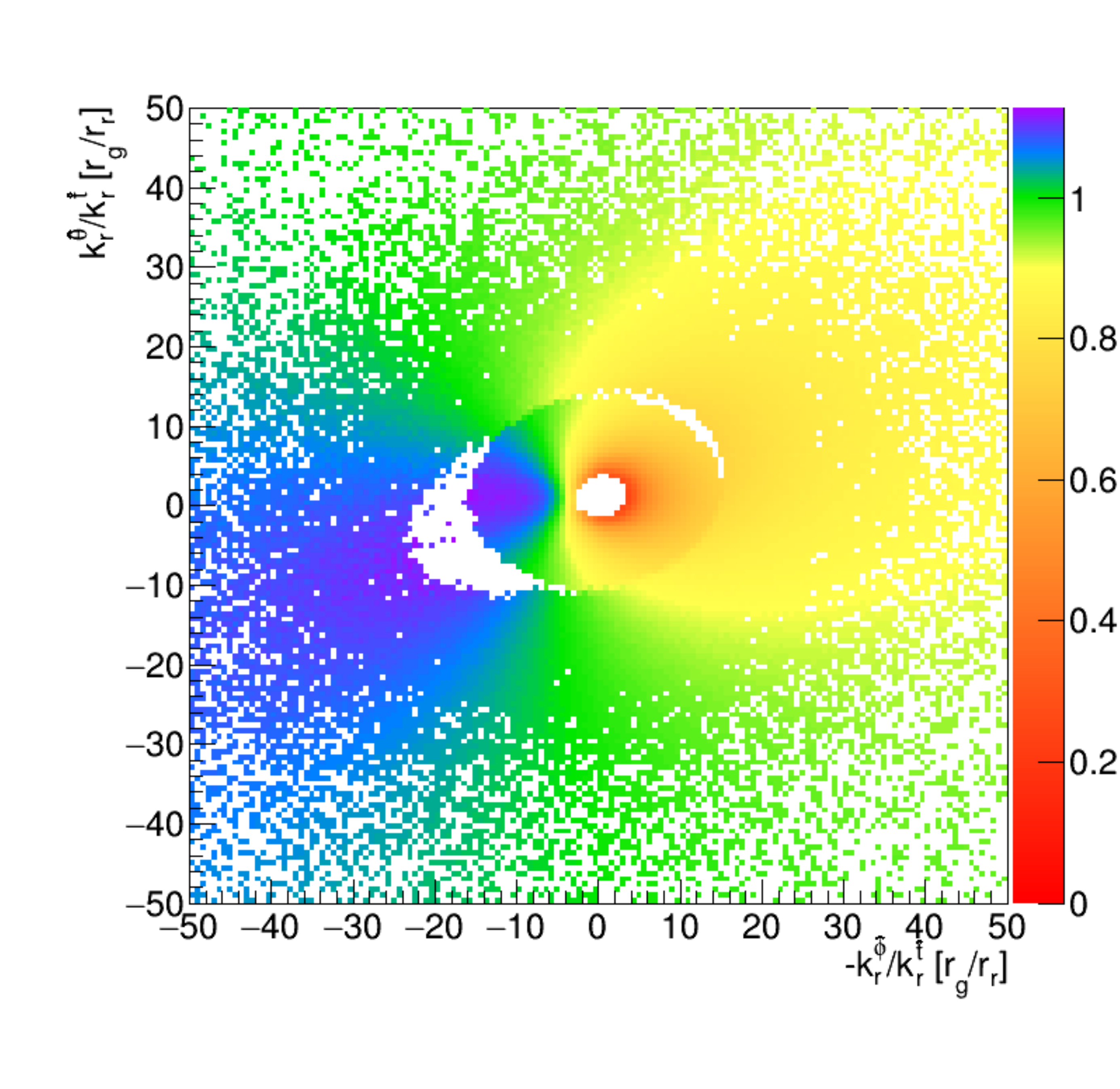}
  \caption{Maps of the frequency shift factor between emission and observation for our default disk configuration. The inclination of the inner disk is fixed at \ang{45}. On the left, the azimuth is \ang{90} and on the right it is \ang{270}.} 
  \label{fig:shadowsT15}
\end{figure*}

\begin{figure*}
  \centering
  \includegraphics[width=0.4\linewidth]{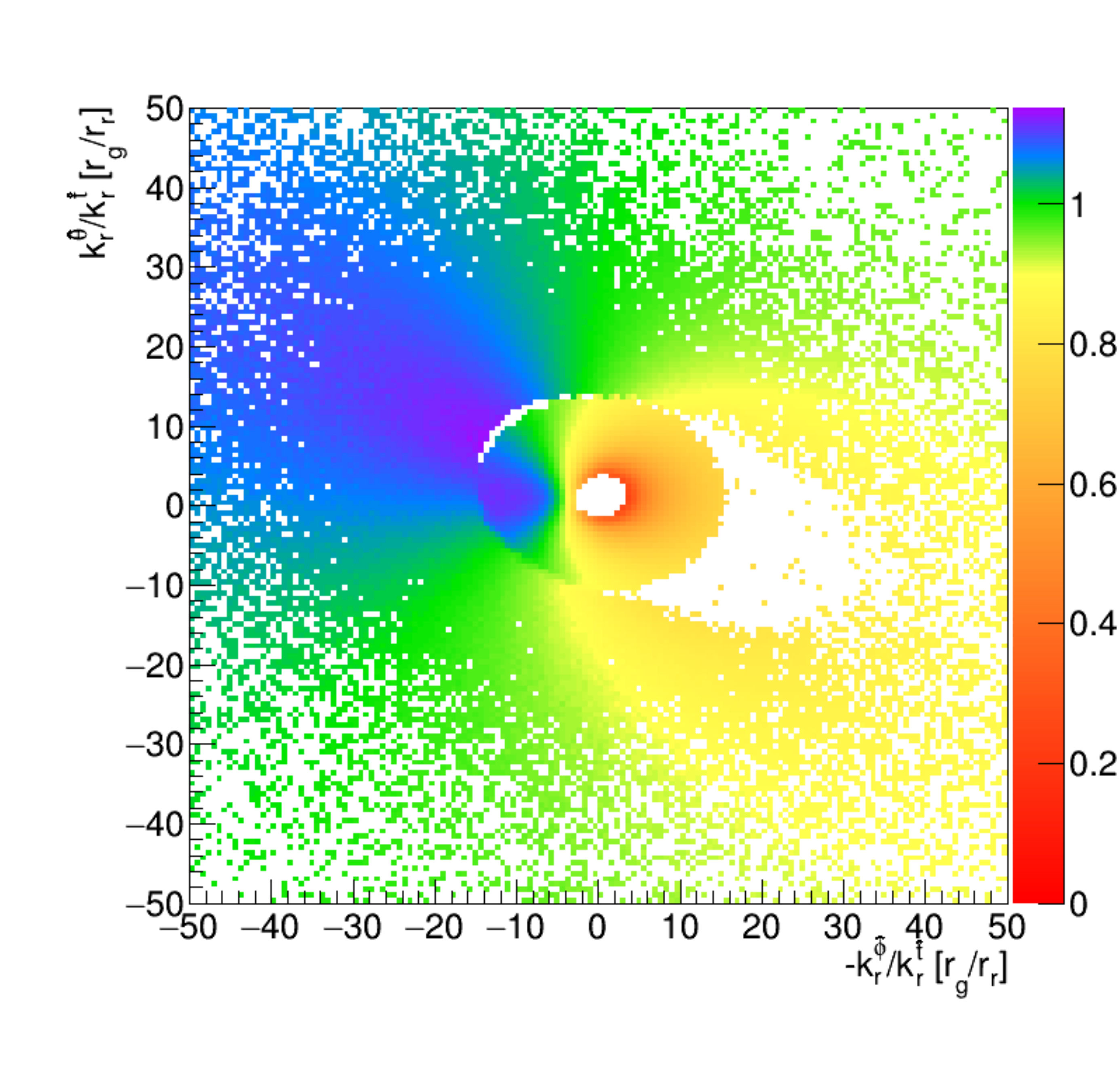}
  \includegraphics[width=0.4\linewidth]{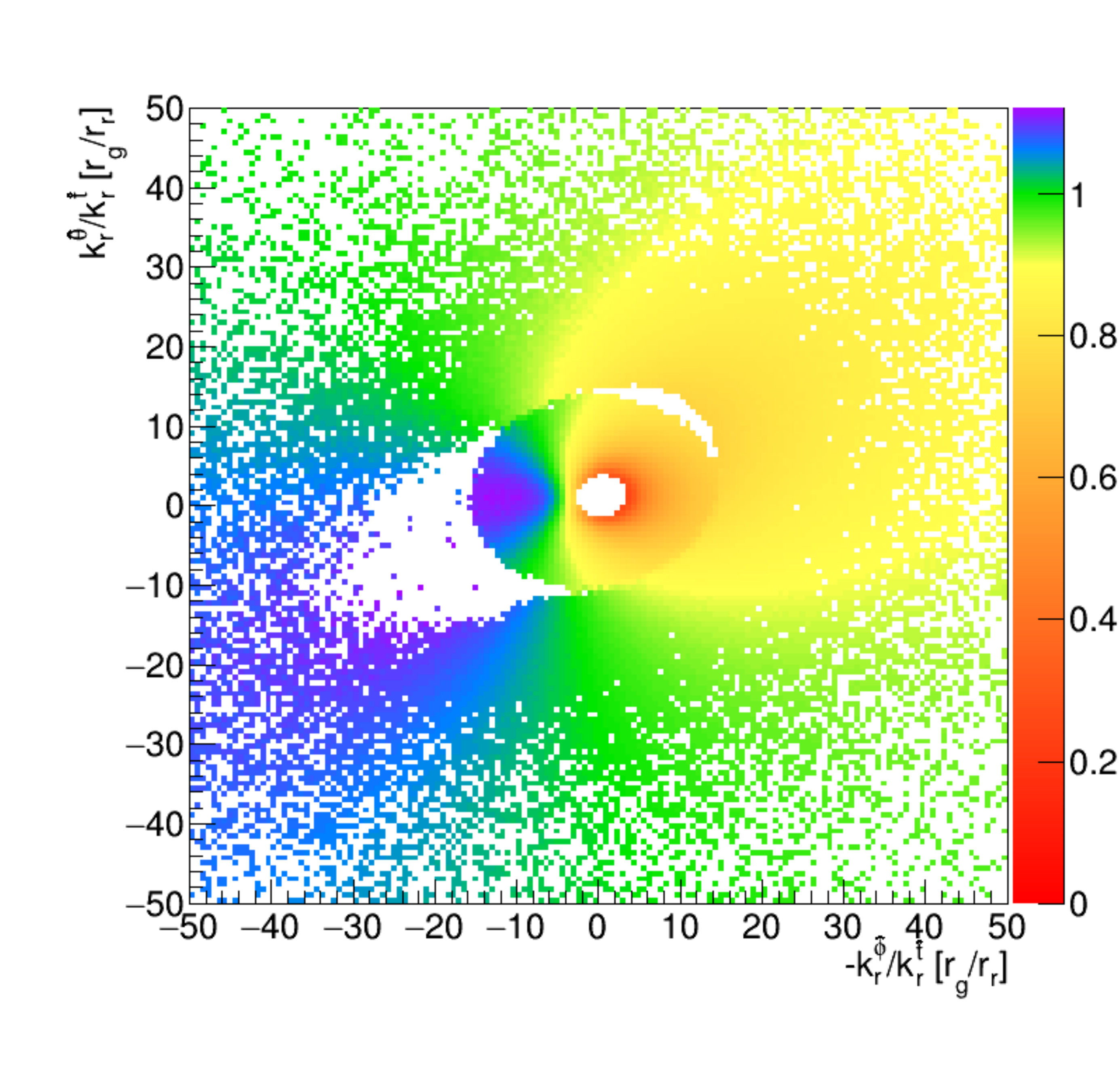}
  \caption{As in Figure \ref{fig:shadowsT15} but for an outer disk tilt of \ang{20}. The shadow cast on the outer disk is quite clear in the redshifted region of the \eq{$\phi$}{\ang{90}} on the left and in the blueshifted region of the \eq{$\phi$}{\ang{270}} on the right.}
  \label{fig:shadowsT20}
\end{figure*}

With forward ray-tracing photons from the lamppost, our code is able to include the effect of shadows cast on the outer disk by the inner disk.
In Figure \ref{fig:shadowsT15} we show two images of the disk for an observer at  \ang{45} inclination, with the color bar indicating the frequency shift of photons.
The shadow is seen by the gap in the portion of the outer disk below the equatorial plane, and is not very large.
Increasing the outer disk tilt to \ang{20} makes the shadow more significant, as shown in Figure \ref{fig:shadowsT20}.
The observer's azimuth detemines whether the redshifted or blueshifted portion of the disk will be shadowed: the red region at \eq{$\phi$}{\ang{90}}, and blue at \eq{$\phi$}{\ang{270}}.
As we vary the observer azimuth, then, not only will the contribution from the outer disk shift in energy as the outer disk changes inclination, but also the red and blue horns of this contribution will be supressed as the corresponding portion of the disk is shadowed.


\subsection{The Line Profile from Warped Disks} \label{sec:line}
\begin{figure*}
    \centering
    \includegraphics[width=0.9\linewidth]{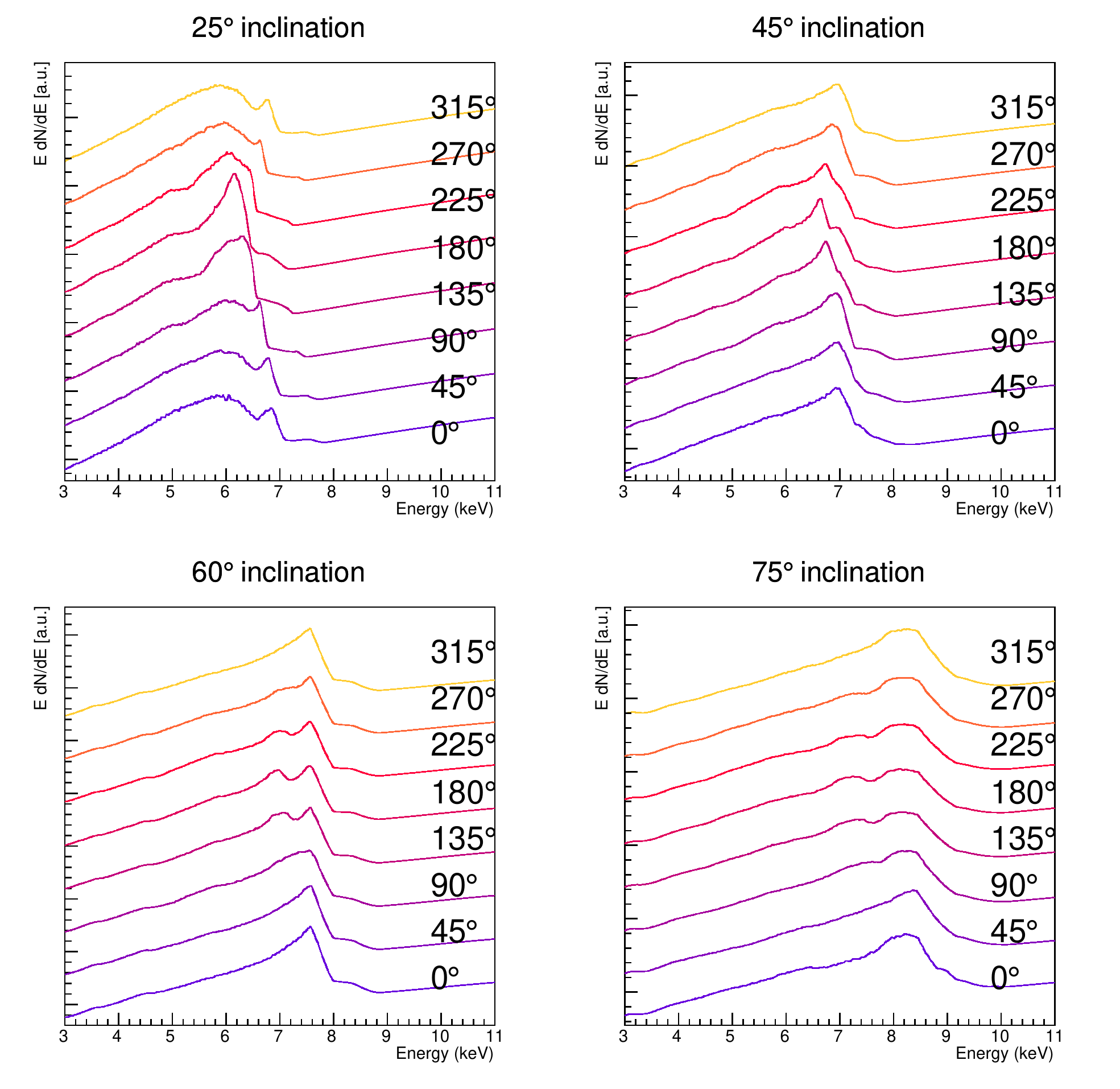}
    \caption{Line profiles seen by observers at different azimuth angles. The profiles are normalized to the same energy flux above \SI{6.4}{\kev} and have been offset from one another for clarity. Each profile is labeled with the observer's azimuth on the right hand side.}
    \label{fig:profile_az}
\end{figure*}

In Figure \ref{fig:profile_az}, we look at the iron line profiles for observers at four inclinations between \eq{$\theta$}{\ang{25}} and \eq{$\theta$}{\ang{75}} at azimuths all around the system.
For the small \rbp{} value used here, the reflection off the outer disk contributes significantly to the shape of the profile.

The tendency of a warped disk seems to be to smear out the iron line further, though if the inclinations are different enough two peaks may be visible.
This tends to occur around \eq{$\phi$}{\ang{180}}, where the outer disk is seen with its lowest inclination relative to the inner disk, and thus the blue horn contributed to the line profile is at its lowest energy.
This is most obvious in the \ang{60} inclination plot (lower left) of Figure \ref{fig:profile_az}.
The contribution from the inner disk is peaked around \SI{7.5}{\kev}.
Around \eq{$\phi$}{\ang{0}}, the outer disk is inclined at about \ang{75}, so contributes less flux at higher energies.
As the azimuth approaches \ang{180}, the outer disk inclination approaches \ang{45} and the blue horn shifts down in energy to just below \SI{7}{\kev}, becoming clearly separate from the inner disk contribution.
Though in our simulated data these two peaks are also distinct in the bottom right plot of \ang{75} inclination, the higher inclination means that both peaks are already smeared and are thus less sharply separated.

Looking at the profiles of \eq{$\phi$}{\ang{90}} and \eq{$\phi$}{\ang{270}}, we might naively expect them to be the same: their outer disks have the same inclination and are simply rotated with respect to the inner disk.
The profiles, though, show that the blueshifted contribution from the outer disk is  less prominent  at \eq{$\phi$}{\ang{270}} -- the region of the outer disk orbiting away from the observer is elevated above the inner disk and thus is irradiated more by the lamppost, while the blue region is irradiated less as well as shadowed.

\begin{figure*}[h]
    \centering
    \includegraphics[width=0.32\linewidth]{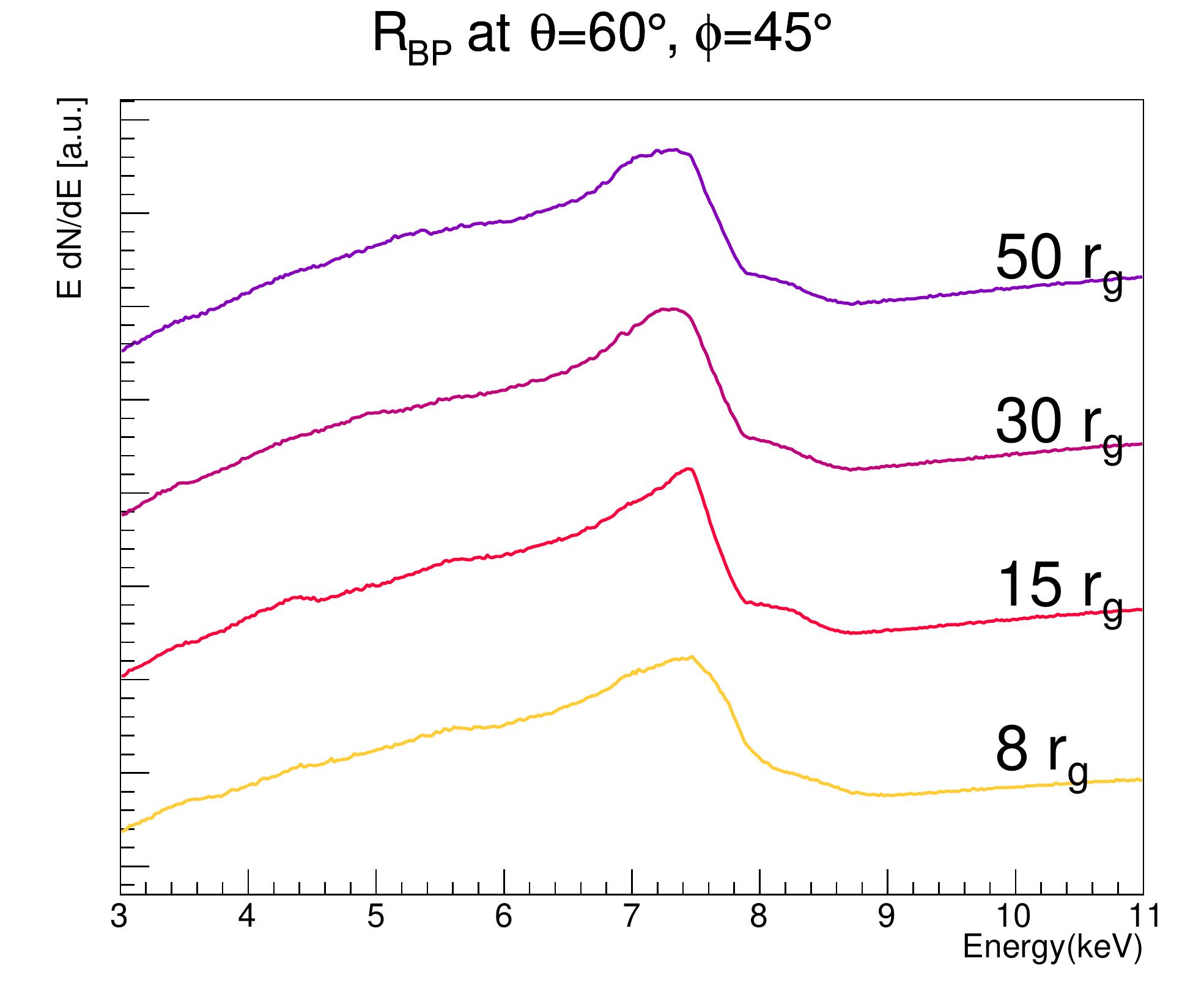}
    \includegraphics[width=0.32\linewidth]{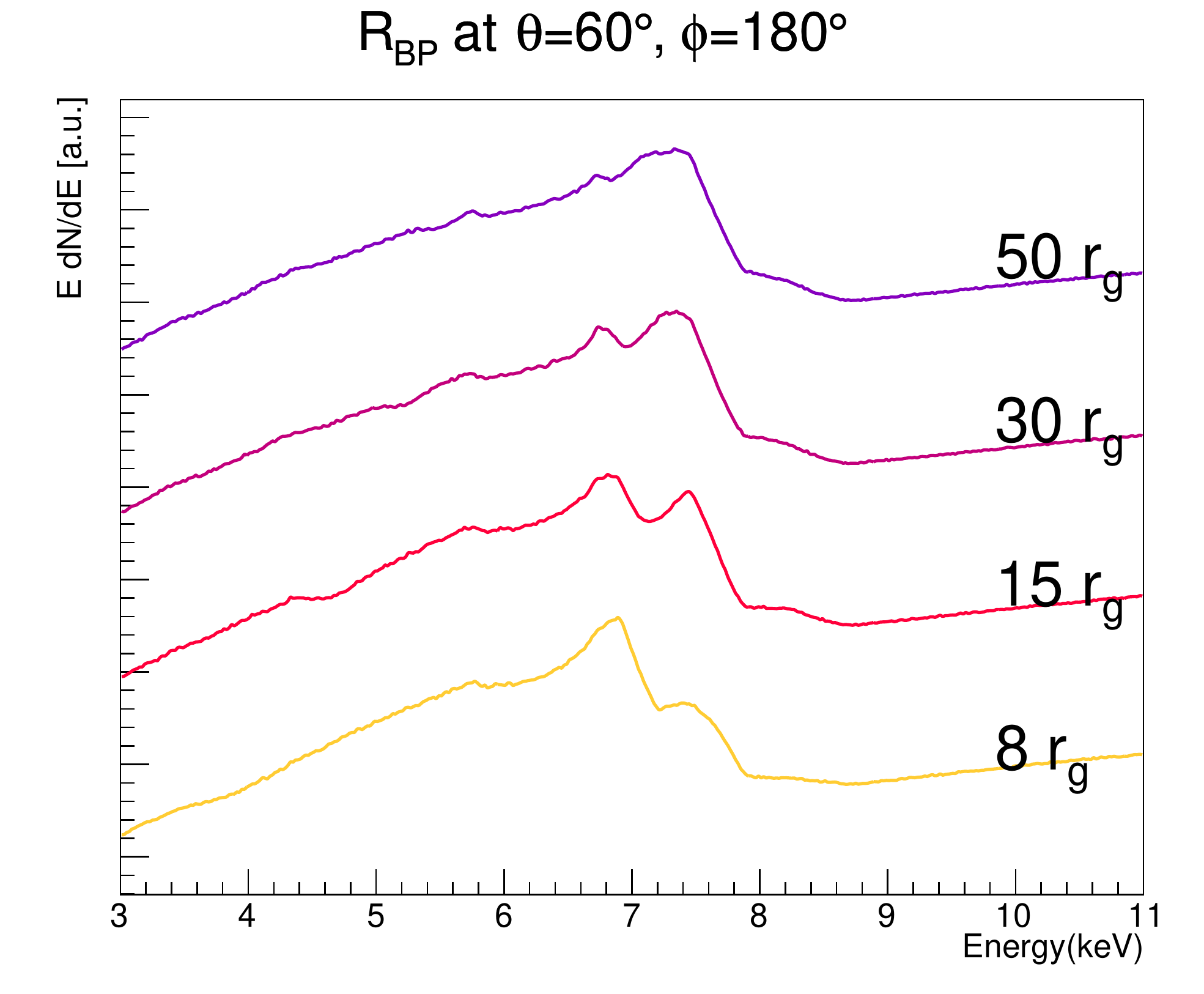}
    \includegraphics[width=0.32\linewidth]{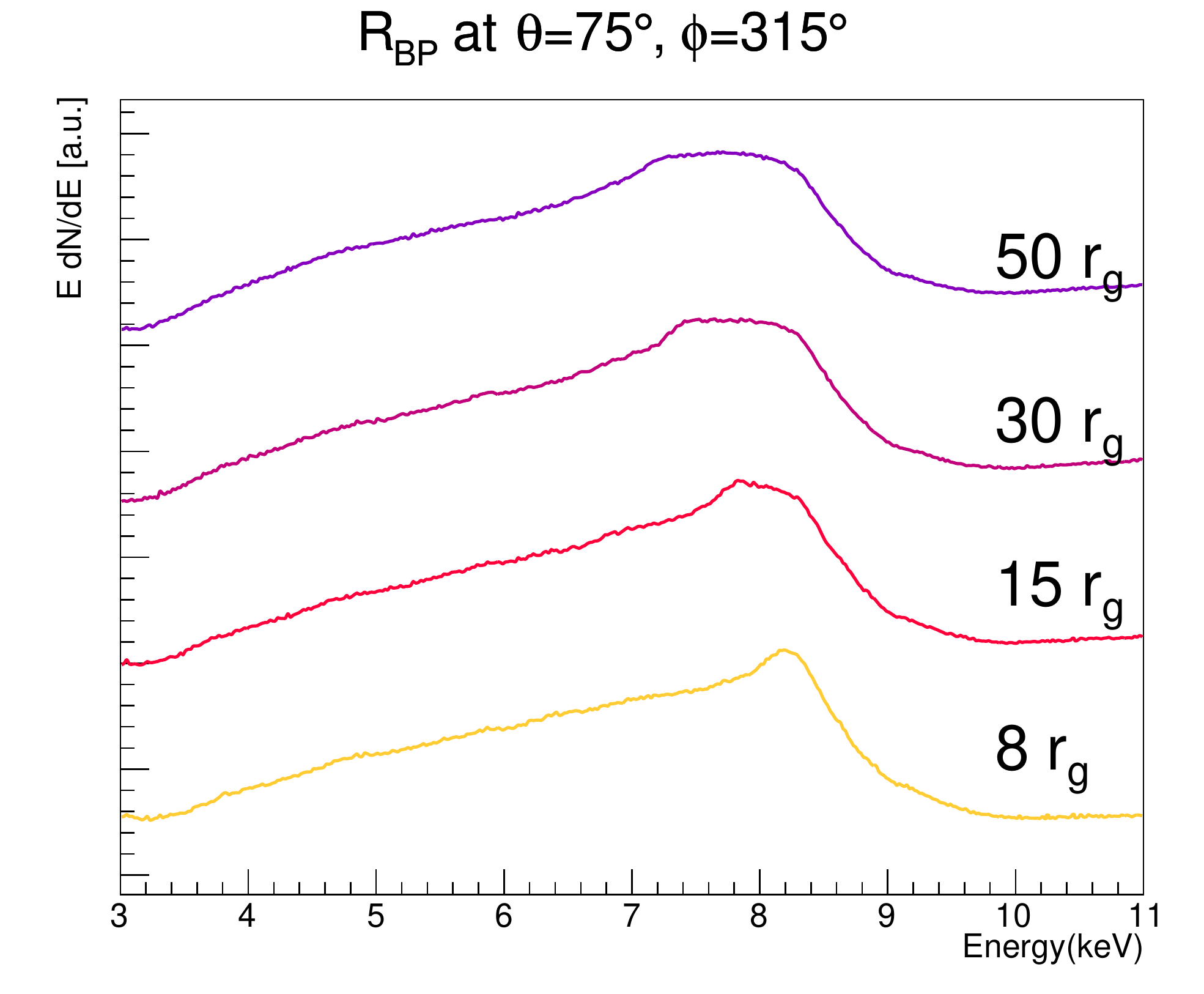}
    \caption{Line profiles for four different values of \rbp{} between \SIrange[range-phrase={ and }]{8}{50}{\rg}, labeled on the right. The other simulation parameters are fixed, with \eq{$a$}{0.9}, \eq{$\beta$}{\ang{15}}, and \eq{\h{}}{\SI{5}{\rg}}. The left plot is seen by the observer at \eq{$\theta$}{\ang{60}} and \eq{$\phi$}{\ang{45}}, the middle plot is at \eq{$\theta$}{\ang{60}} and \eq{$\phi$}{\ang{180}}, and the right plot is at \eq{$\theta$}{\ang{675}} and \eq{$\phi$}{\ang{315}}.} 
    \label{fig:profile_r}
\end{figure*}

Next, we examine the ways that changing properties of the disk misalignment affects the reflected spectrum.
In Figure \ref{fig:profile_r}, we show the profile for four values of \rbp{}, between \SIrange[range-phrase= { and }, range-units=single]{8}{50}{\rg}, for several different observers.
These three viewing angles exhibit the most salient features seen in all the simulations.
We vary the warp radius while holding the tilt and lamppost height constant at \ang{15} and \SI{5}{\rg}, respectively.

\begin{figure*}[h]
    \centering
    \includegraphics[width=0.32\linewidth]{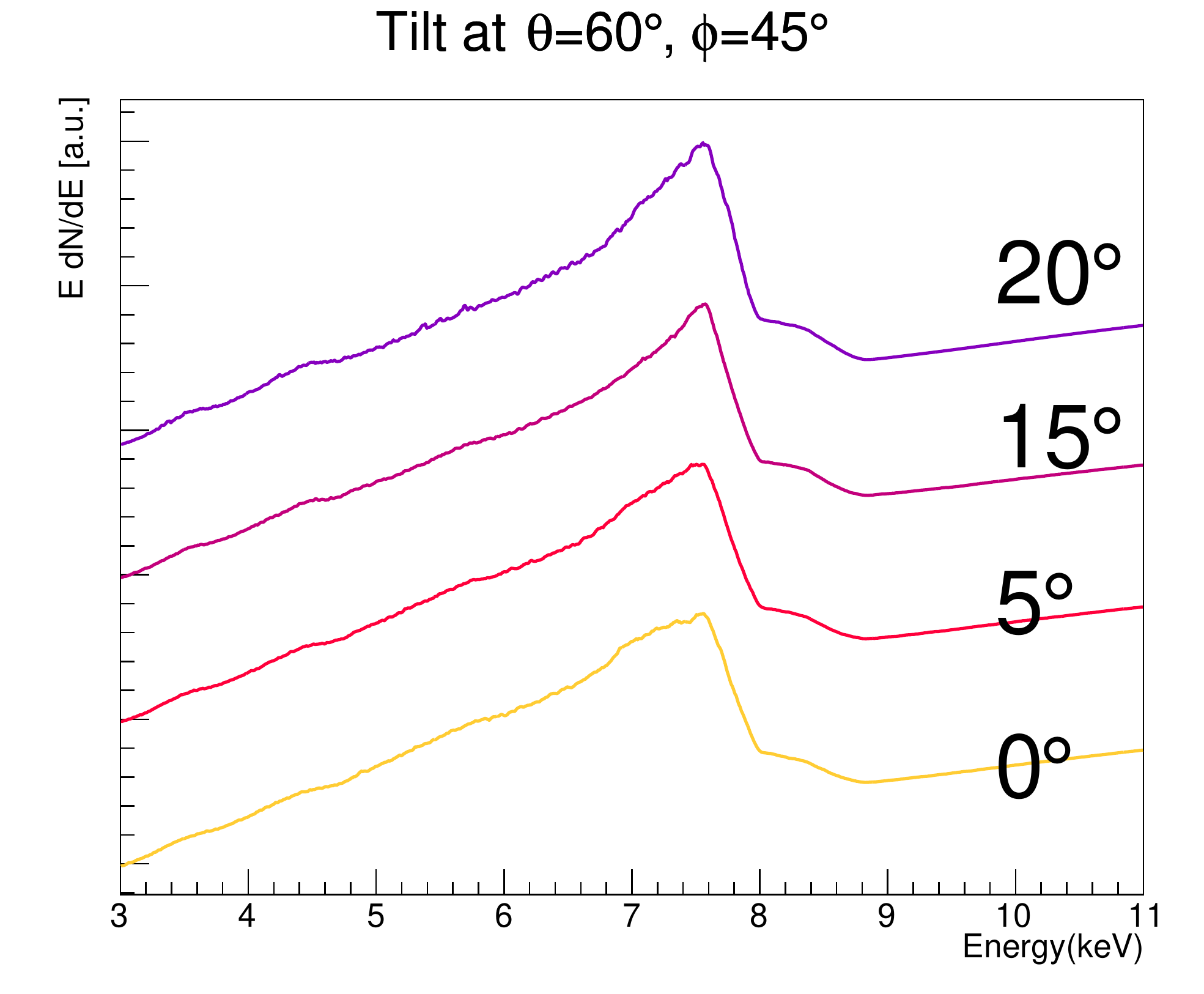}
    \includegraphics[width=0.32\linewidth]{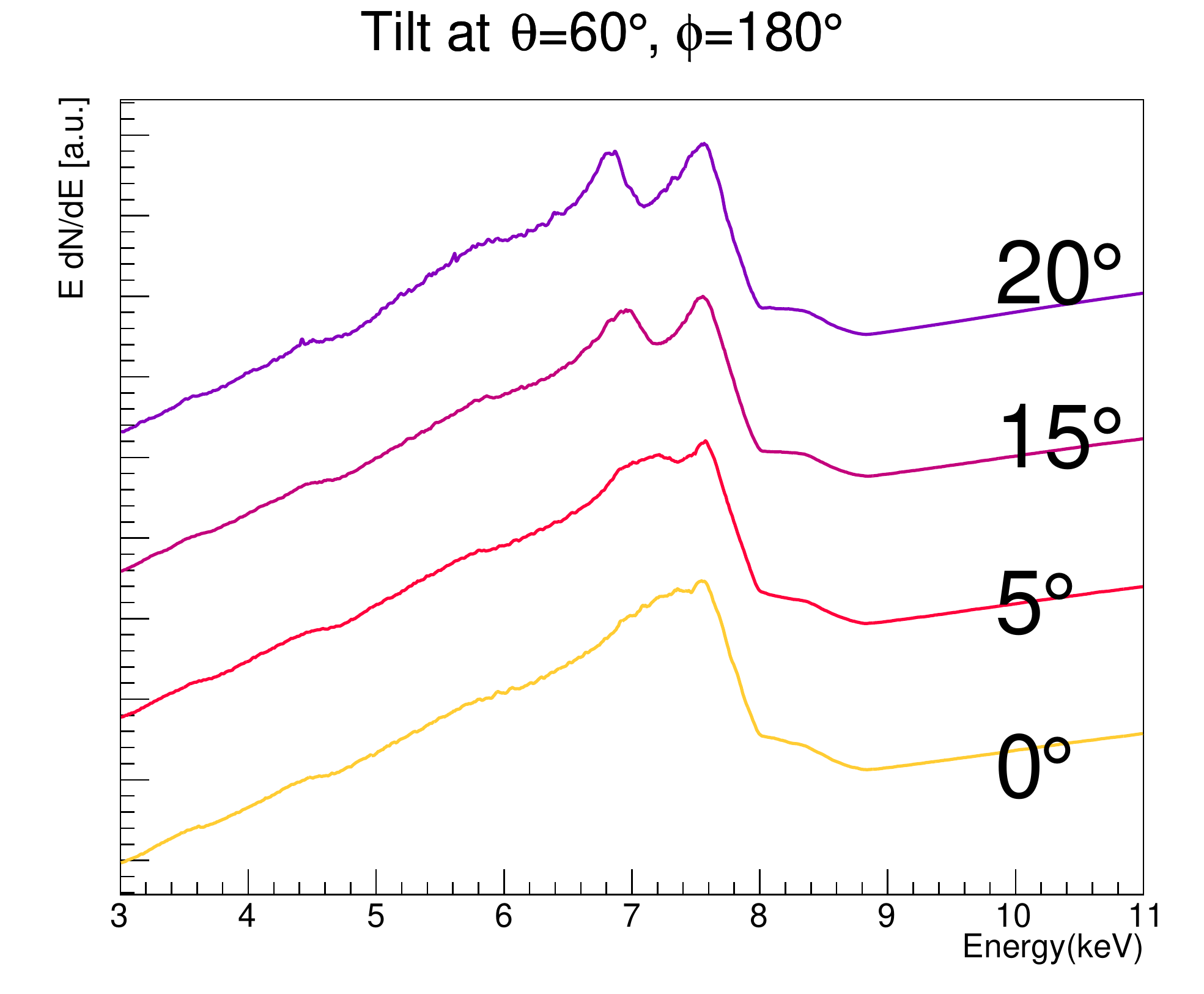}
    \includegraphics[width=0.32\linewidth]{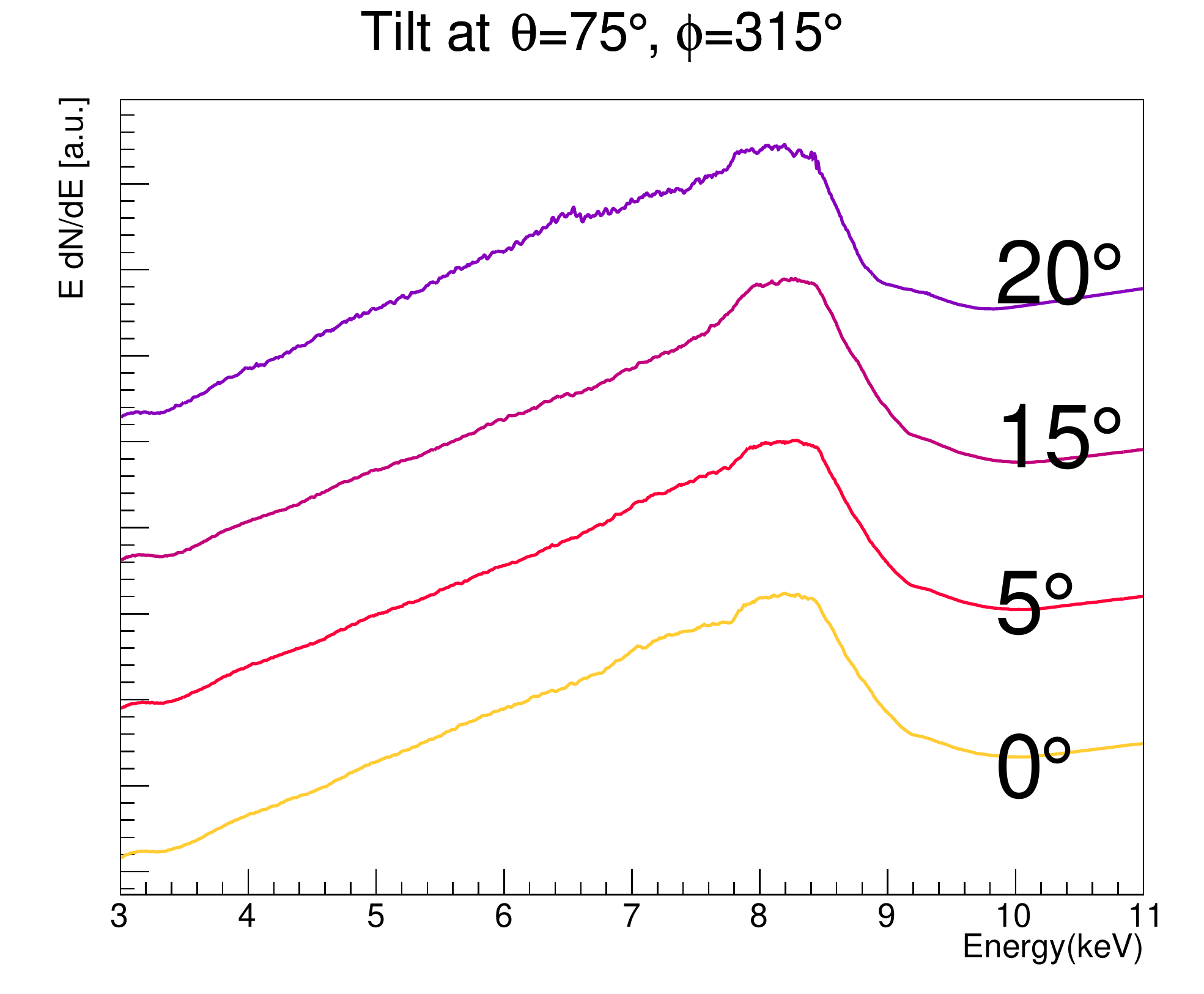}
    \caption{Line profiles for four different values of $\beta$ between \SIrange[range-phrase={ and }]{0}{20}{\degree}, labeled on the right. The other simulation parameters are fixed, with \eq{$a$}{0.9}, \eq{\rbp{}}{\SI{15}{\rg}}, and \eq{\h{}}{\SI{5}{\rg}}. The plots are for the same observers as in Figure \ref{fig:profile_r}.}
    \label{fig:profile_beta}
\end{figure*}

\begin{figure*}[h]
    \centering
    \includegraphics[width=0.32\linewidth]{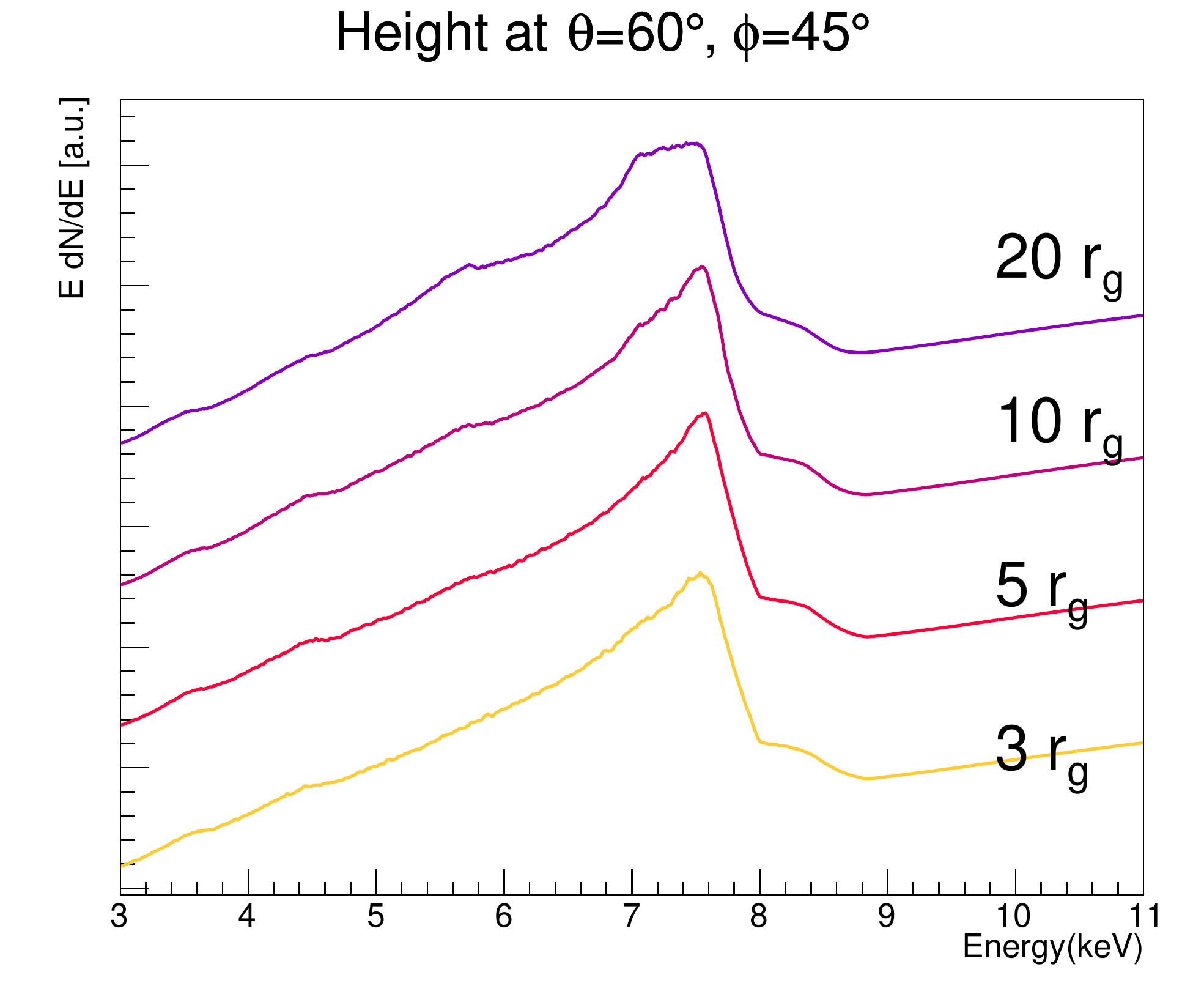}
    \includegraphics[width=0.32\linewidth]{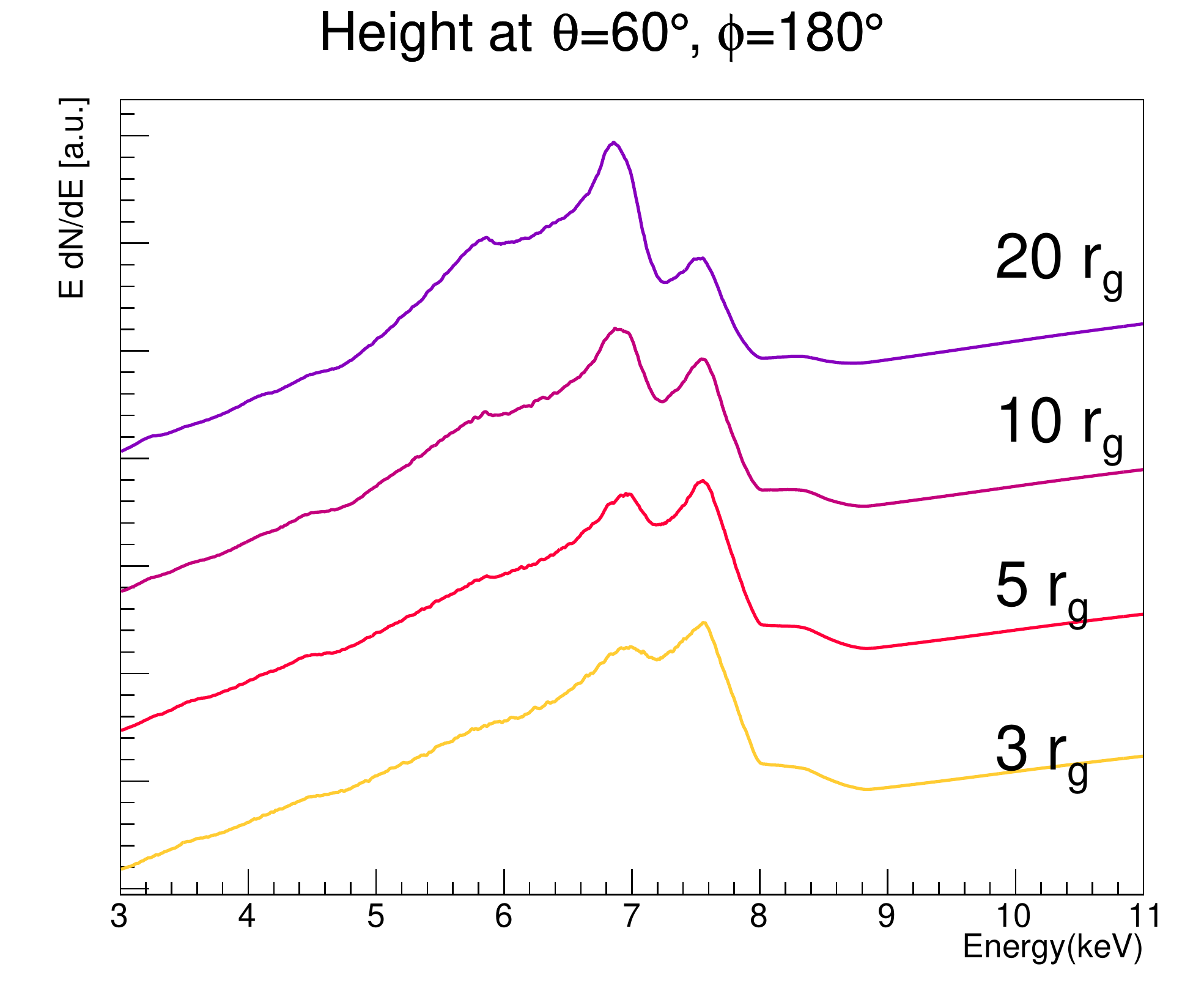}
    \includegraphics[width=0.32\linewidth]{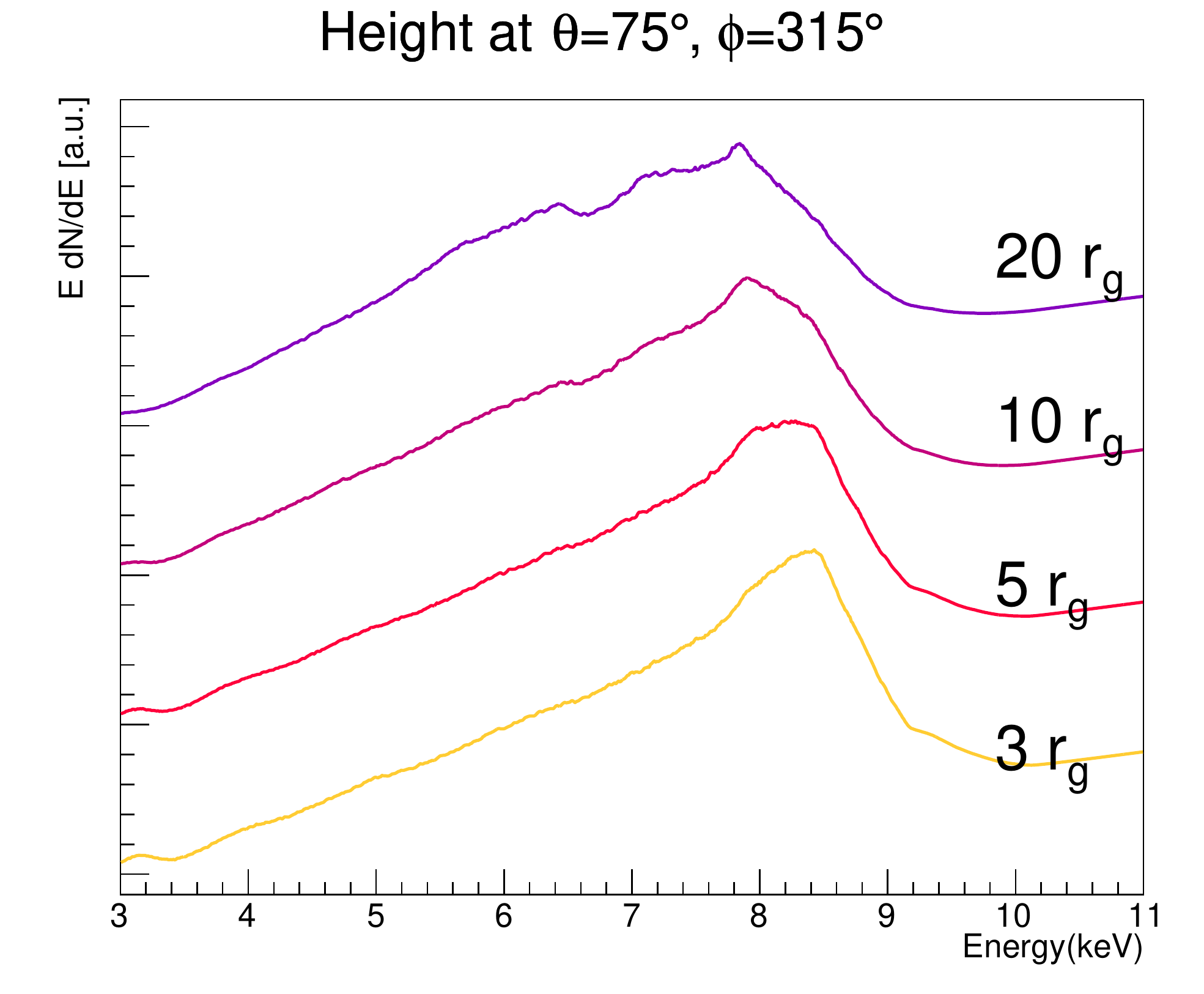}
    \caption{Line profiles for four different values of lamppost height between \SIrange[range-phrase={ and }]{3}{20}{\rg}, labeled on the right. The other simulation parameters are fixed, with \eq{$a$}{0.9}, \eq{\rbp{}}{\SI{15}{\rg}}, and \eq{$\beta$}{\ang{15}}. The plots are for the same observers as in Figure \ref{fig:profile_r}.}
    \label{fig:profile_height}
\end{figure*}

It is clear that the azimuth from which the system is viewed clearly has a strong effect on the reflected spectrum.
At \ang{60} inclination and \ang{45} azimuth, where the inclination of the outer disk is approximately equal to the inclination of the inner disk (\ang{61.12}), the profile is largely unchanged, aside from some minor differences in the shape of the peak.

When the outer disk has a lower inclination than the inner disk, as in the middle at \ang{60} inclination and \ang{180} azimuth (\eq{\iout{}}{\ang{45}}), the two visible peaks from the inner and outer disk trade prominence as the warp radius extends.
Thinking about \rbp{} as the ``ISCO'' of the outer disk, as \rbp{} extends further out this emission peaks at lower and lower energies and contributes less to the total reflected flux.
This causes the energy of the peak flux to shift from being contributed by the outer disk to the inner disk, while in the middle they are relatively equal.

In the right plot, the inner disk inclination is \ang{75}, the azimuth is \ang{315}, and the outer disk is nearly edge on at \eq{\iout{}}{\ang{85.8}}.
Here we do not see two peaks; instead, as the inner edge of the outer disk \rbp{} gets larger, this emission from the outer disk moves to lower energies, causing the peak to smear out.
The falling edge of the Fe K$\alpha$ peak remains in the same place, the plateau of maximum flux extends, apparently pushing the energy of the peak lower; in Section \ref{sec:fitting}, we show that this broadening of the iron line causes the spin to be overestimated.

Figure \ref{fig:profile_beta} shows the profile for different values of the disk tilt seen by the same observers in Figure \ref{fig:profile_r}.
The middle plot, where the outer disk has a lower inclination than the inner disk, shows that increasing disk tilt shifts the blue horn of the outer disk to lower energies -- the higher tilt lowers the inclination of the outer disk even further.
In the other two plots, the increasing tilt shifts the outer disk emission towards higher energies, though it is already obscured by the peak from the inner disk.

Figure \ref{fig:profile_height} shows how the profile changes with lamppost height.
The taller the lamppost is, the larger the break radius of the emissivity, and the more prominent the contribution of the outer disk to the profile.
This is very clear in the middle, where at \eq{H}{\SI{20}{\rg}} the entire profile from the outer disk can clearly be seen on top of the contribution from the inner disk.

\subsection{Emissivity profile} \label{sec:emissivity}

\begin{figure}[h!]
    \centering
    \includegraphics[width=0.49\linewidth]{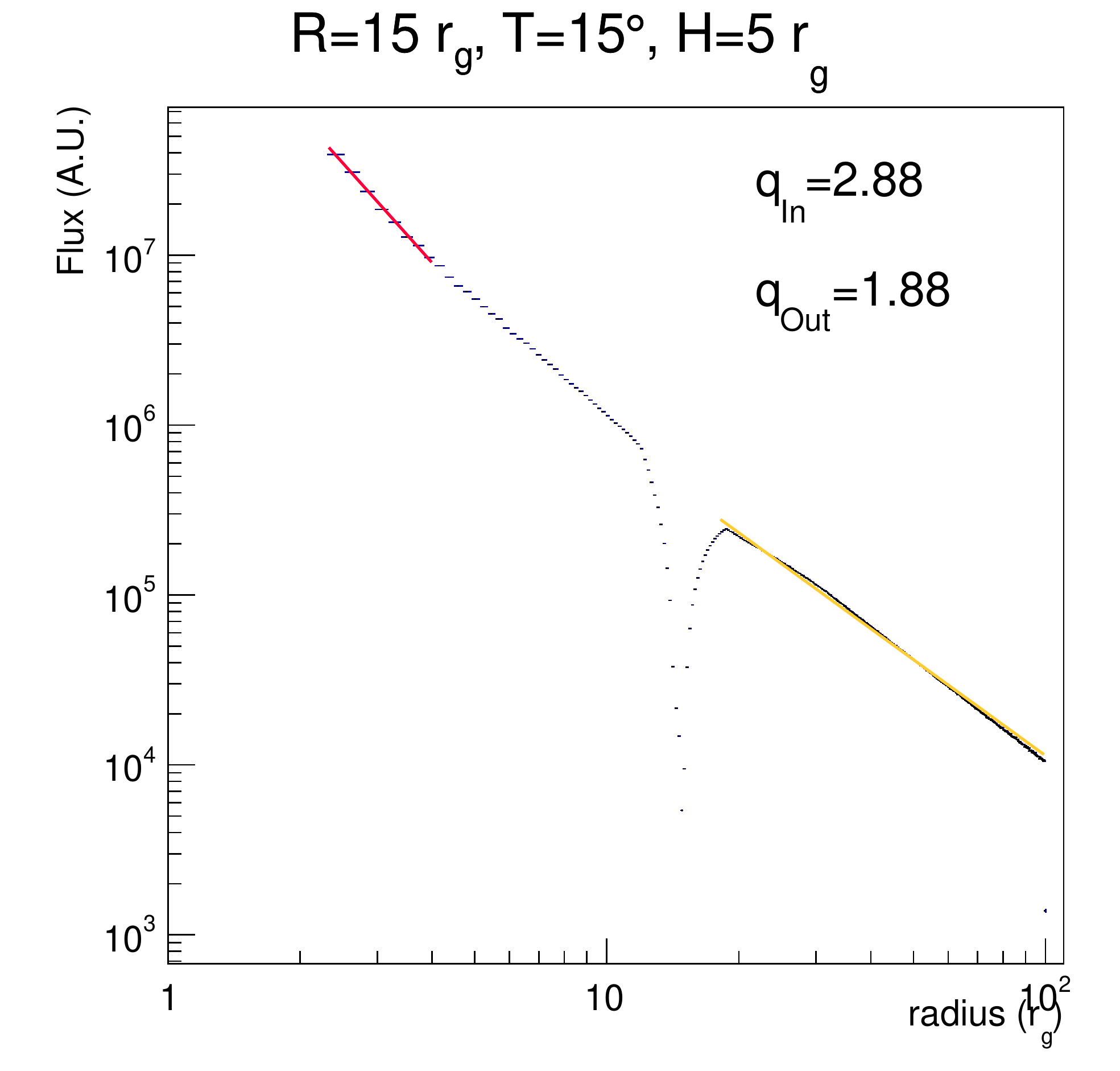}
    \includegraphics[width=0.49\linewidth]{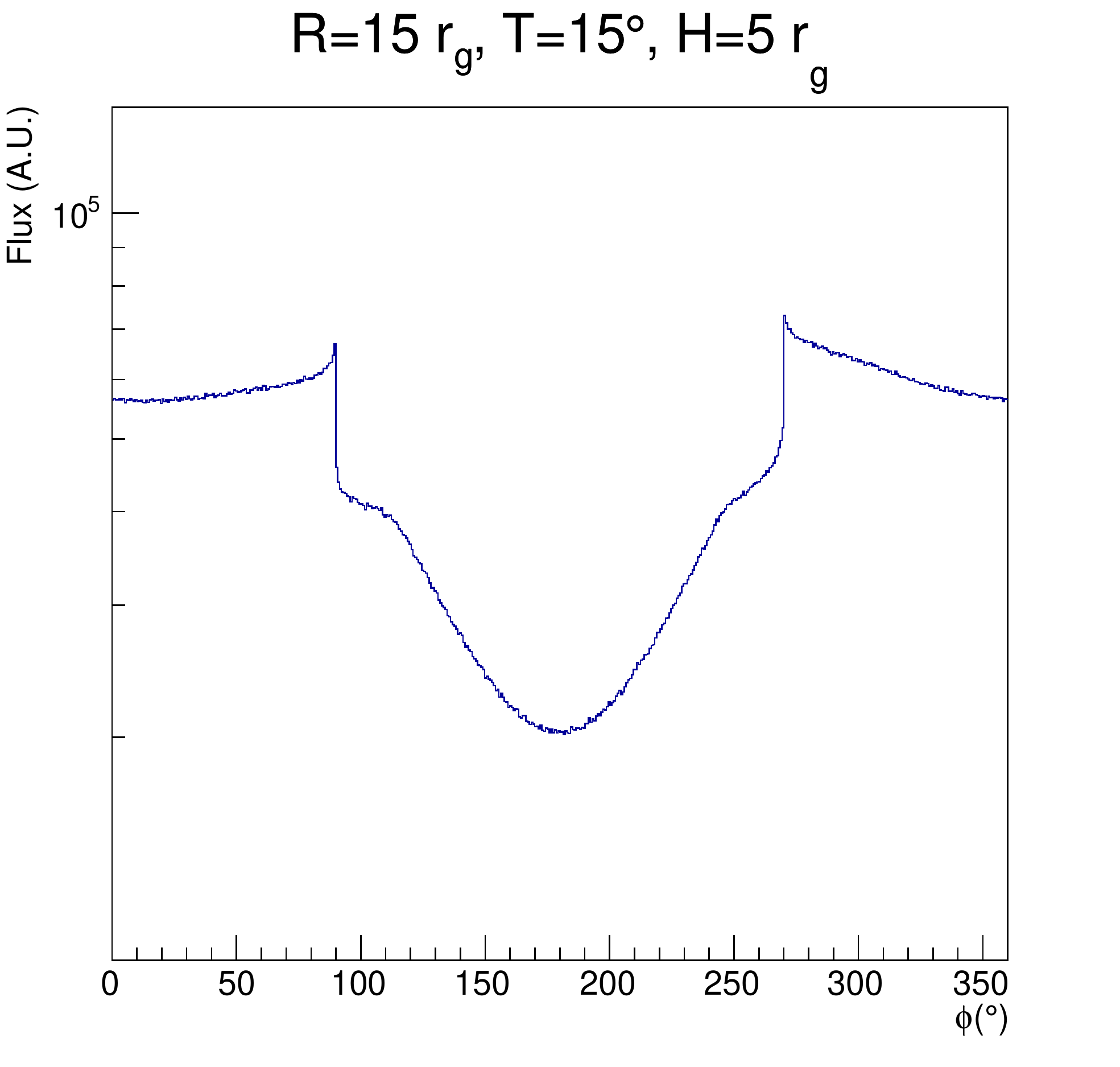}
    \caption{Left: Radially dependent emissivity profile. The emissivity profiles before and after the indentation have been fitted with the indices $q_{\rm in}$ and $q_{\rm out}$. Right: $\phi$-dependent profile for the outer disk (\SI{15}{\rg}$<r<$\SI{100}{\rg}), highlighting the asymmetry caused by the lamppost being off-axis as well as the effects of shadowing the outer disk by the inner disk.}
    \label{fig:em0}
\end{figure}

Figure \ref{fig:em0} shows the emissivity profile (i.e.\ the energy flux $\varepsilon$ of photons with an energy exceeding the energy threshold for iron line emission) for the entire disk, as well as the $\phi$-dependent profile for just the outer disk (in this case, \rbp{}\SI{>15}{\rg}).

There is a dip in the profile at \rbp{}, so for fitting the break in emissivity, we fit $q_{in}$ inside \eq{$r$}{\h{}} or \rbp{}, whichever is smaller, (from $\varepsilon\propto r^{-q_{in}}$) and $q_{out}$ outside \rbp{} or \h{}, whichever is larger.
For all simulated models, the results are collected in Table \ref{tab:em}.

\begin{table}[h]
    \renewcommand{\arraystretch}{1.4}
    \centering
    \begin{tabular*}{\linewidth}{c @{\extracolsep{\fill}} c c  c c}
        \hline
        \rbp{}(\SI{}{\rg}) & $\beta$(\SI{}{\degree}) & \h{}(\SI{}{\rg})& q$_{\rm In}$ & q$_{\rm Out}$ \\
        \hline
        15 & 15 & 5 & 2.88 & 1.88 \\
        \hline
        8 & \multirow{3}{*}{15} & \multirow{3}{*}{5} & 2.92 & 2.00 \\
        30 & & & 2.86 & 1.88 \\
        50 & & & 2.86 & 1.90 \\
        \hline
        \multirow{3}{*}{15} & 0 & \multirow{3}{*}{5} & 2.86 & 1.98 \\
        & 5  & & 2.87 & 1.99 \\
        & 20 & & 2.89 & 1.83 \\
        \hline
        \multirow{3}{*}{15} & \multirow{3}{*}{15} & 3 & 2.74 & 1.83 \\
        & & 10 & 1.76 & 1.92 \\
        & & 20 & 1.21 & 1.75 \\
        \hline
    \end{tabular*}
    \caption{Emissivity indices for inner and outer disks. We show results for our default parameter values, and group the remaining results into three groups for the varied parameter: transition radius \rbp{}, tilt $\beta$, and lamppost height \h{}.}
    \label{tab:em}
\end{table}

\section{Fitting of Simulated Iron Profiles} \label{sec:fitting}

To determine how disk warping effects the spin measurement of a BH, we have imported our profiles into \xspec{} \citep{xspec} and used them to generate synthetic energy spectra.
To make our profiles compatible for fitting with \relxilllp{}, the lamppost version of \relxill{} \citep{garcia2014}, we only use the profiles from photon beams which scatter a single time.
This decision is motivated by the discussion in Section \ref{sec:multiple}; we note there that there is as of yet not publicly available method to produce the \xillver{} contributions from photon beams which scatter multiple times, as the incident spectrum would be a combination of the power law and fluorescent emission from the first scattering.
Based on Figure \ref{fig:profile_scatterings} this reprocessed radiation does not dominate the iron line profile and so can be ignored in this work.
It will be a future avenue of study, supported by the recent work of \cite{wilkins2020} into simulating reprocessed radiation.

To create synthetic spectra from our profiles, we use the \textsc{fakeit} command in \xspec{}, which multiplies the profiles with the response curve from the \textit{Chandra}-HETG Cycle 22 sample response files.
These synthetic profiles have a resolution of \SI{10}{\electronvolt}, similar to that of the microcalorimeter on the upcoming {\it XRISM} \citep{xrism} and {\it Athena} \citep{athena} missions; we expect these results will prove more useful once these two large effective area instruments are in operation.
For our \relxilllp{} model, we fix the inner and outer radius of the disk to the ISCO and \SI{100}{\rg}, respectively; the cutoff energy to \SI{300}{\kev}; the reflected fraction to -1 (so our model doesn't include any of the direct emission from the lamppost); and the redshift of the source to 0.
We freely fit spin, inclination, lamppost height, power law index, the ionization, and the iron abundance, with each initially set to their simulated value.

To verify that our code broadens the reflected energy spectra from \xillver{} in a manner compatible with \relxill{}, we fit our results for a flat, unwarped disk.
We find that the results are relatively accurate, though they are not always within the \SI{90}{\percent} errors.
The inclination can be under- or over-estimated, by up to a degree.
Similarly, the spin can be under/over-estimated by up to \num{0.1}.
We use these results as the standard for how well the warped disk results are fit by the \relxilllp{} model.
In Appendix \ref{ap:fits}, we include the full results of results of fitting the unwarped disk.

\begin{figure*}
    \centering
    \includegraphics[width=0.48\linewidth]{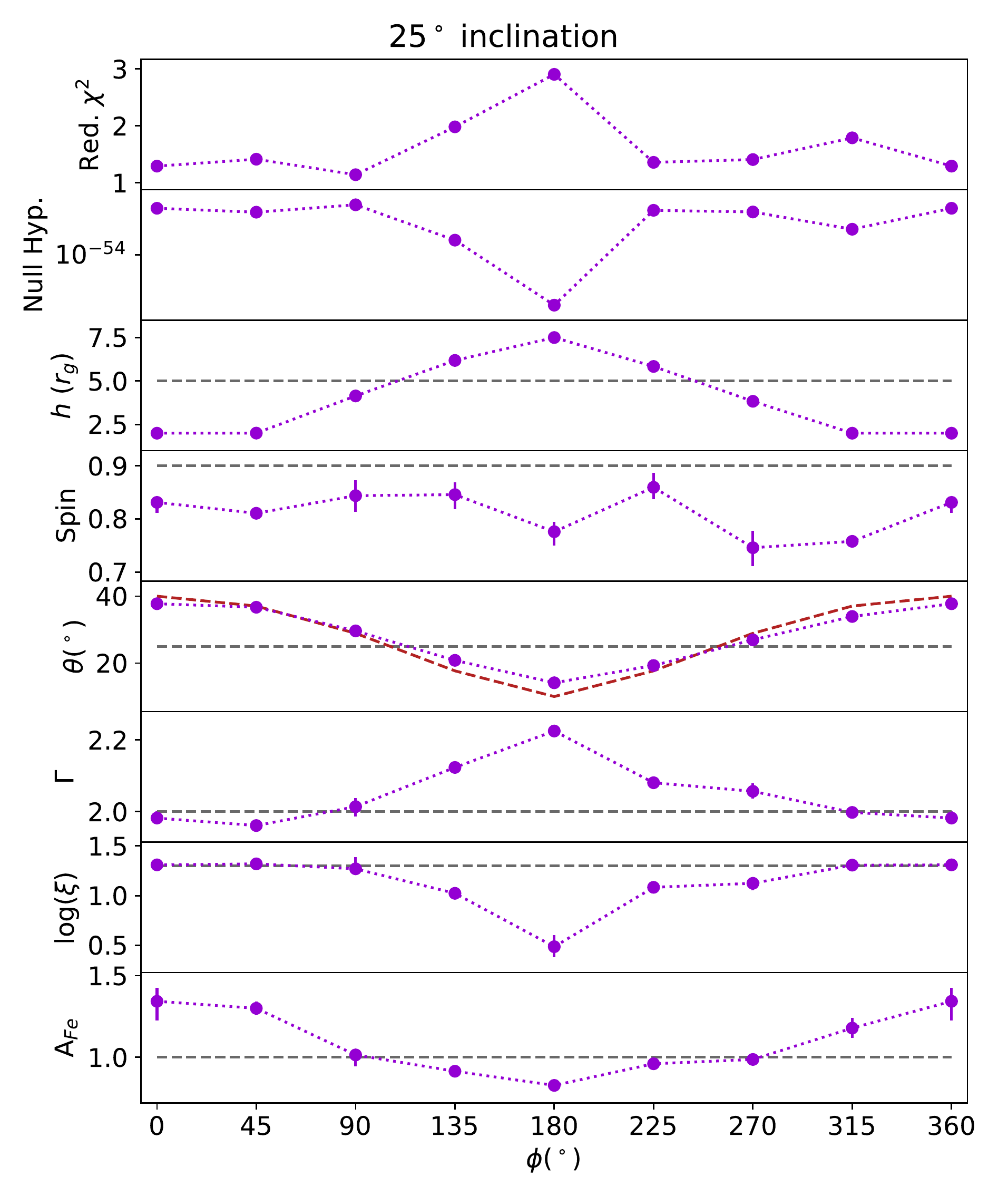}
    \includegraphics[width=0.48\linewidth]{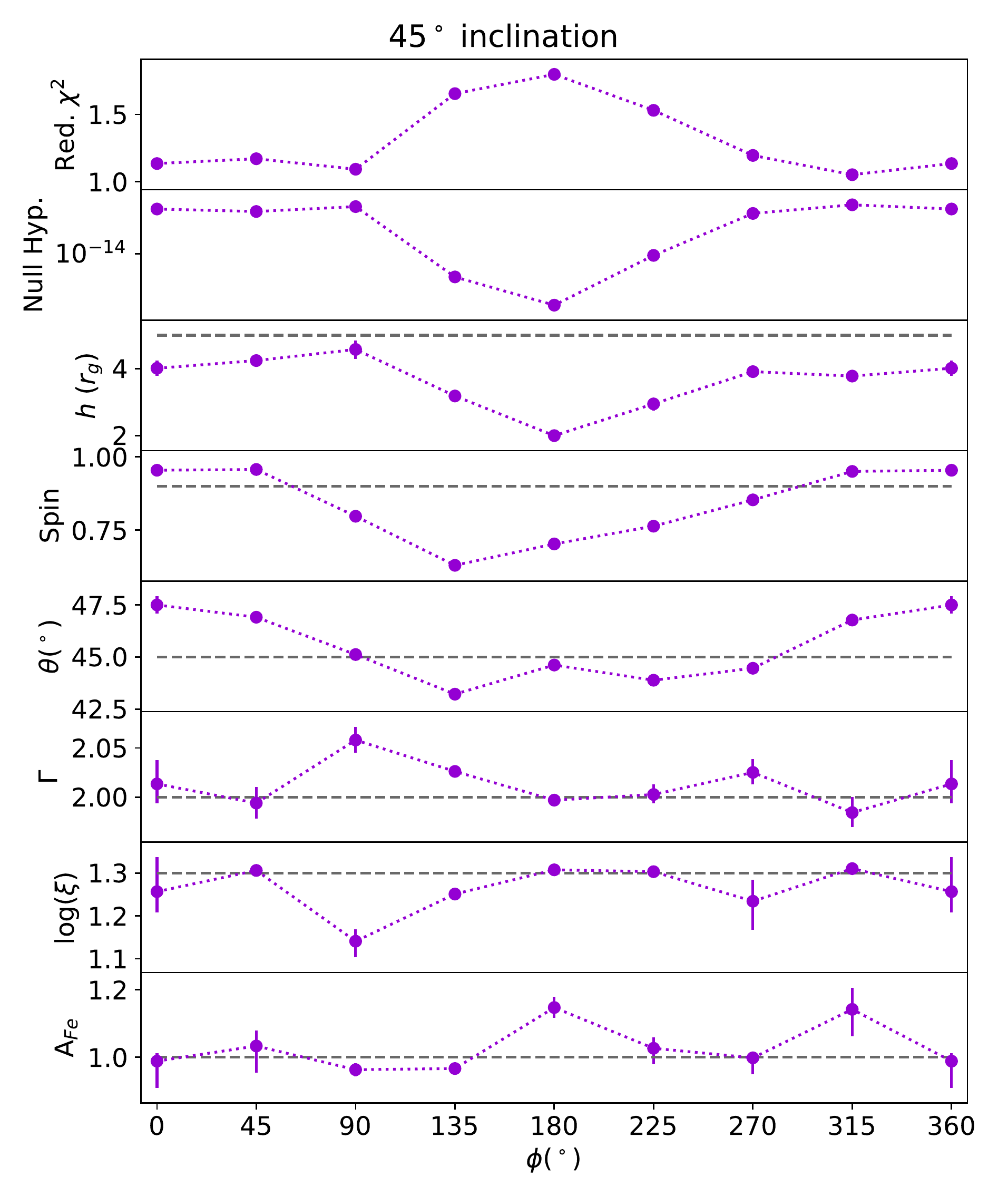}\\
    \includegraphics[width=0.48\linewidth]{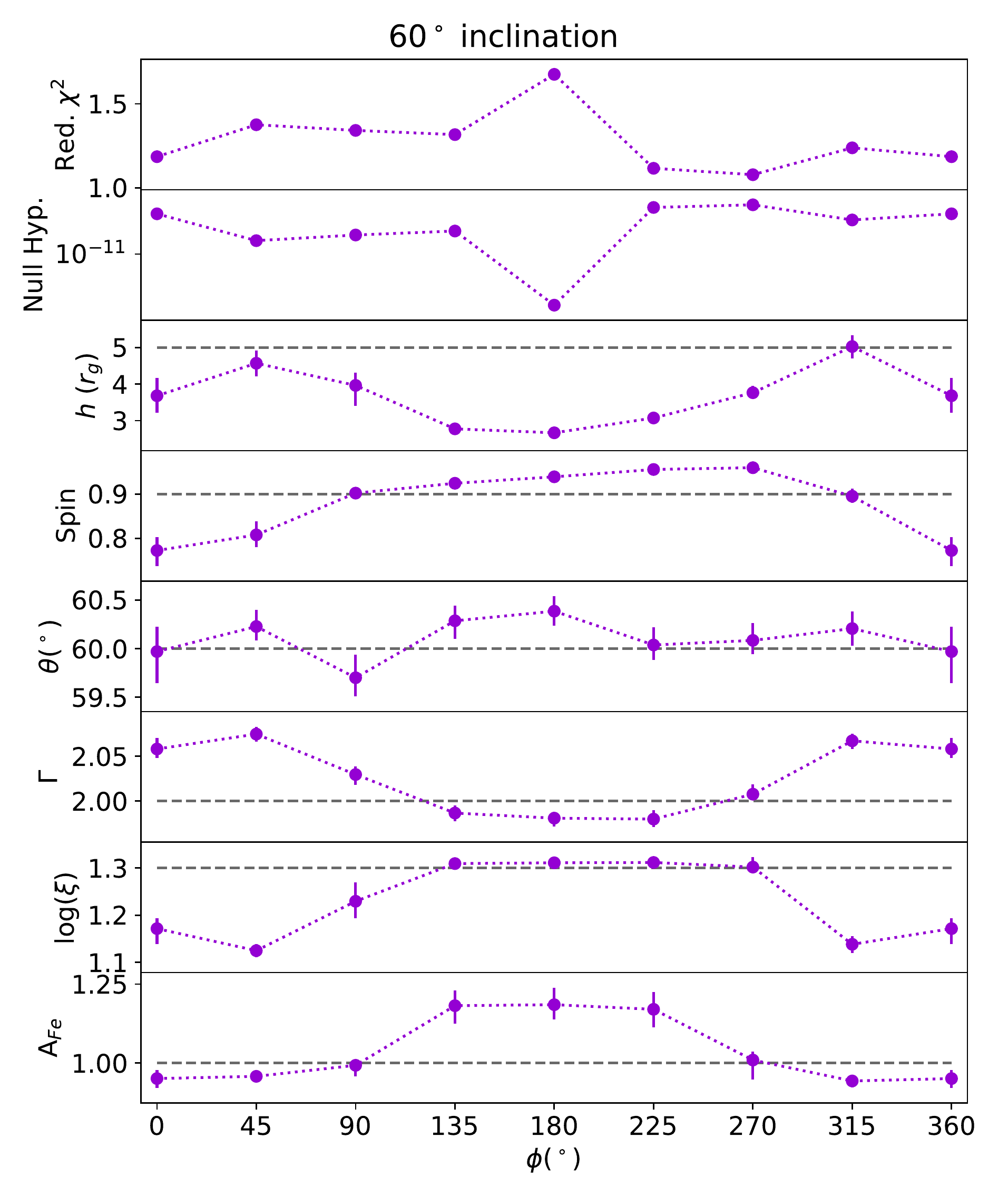}
    \includegraphics[width=0.48\linewidth]{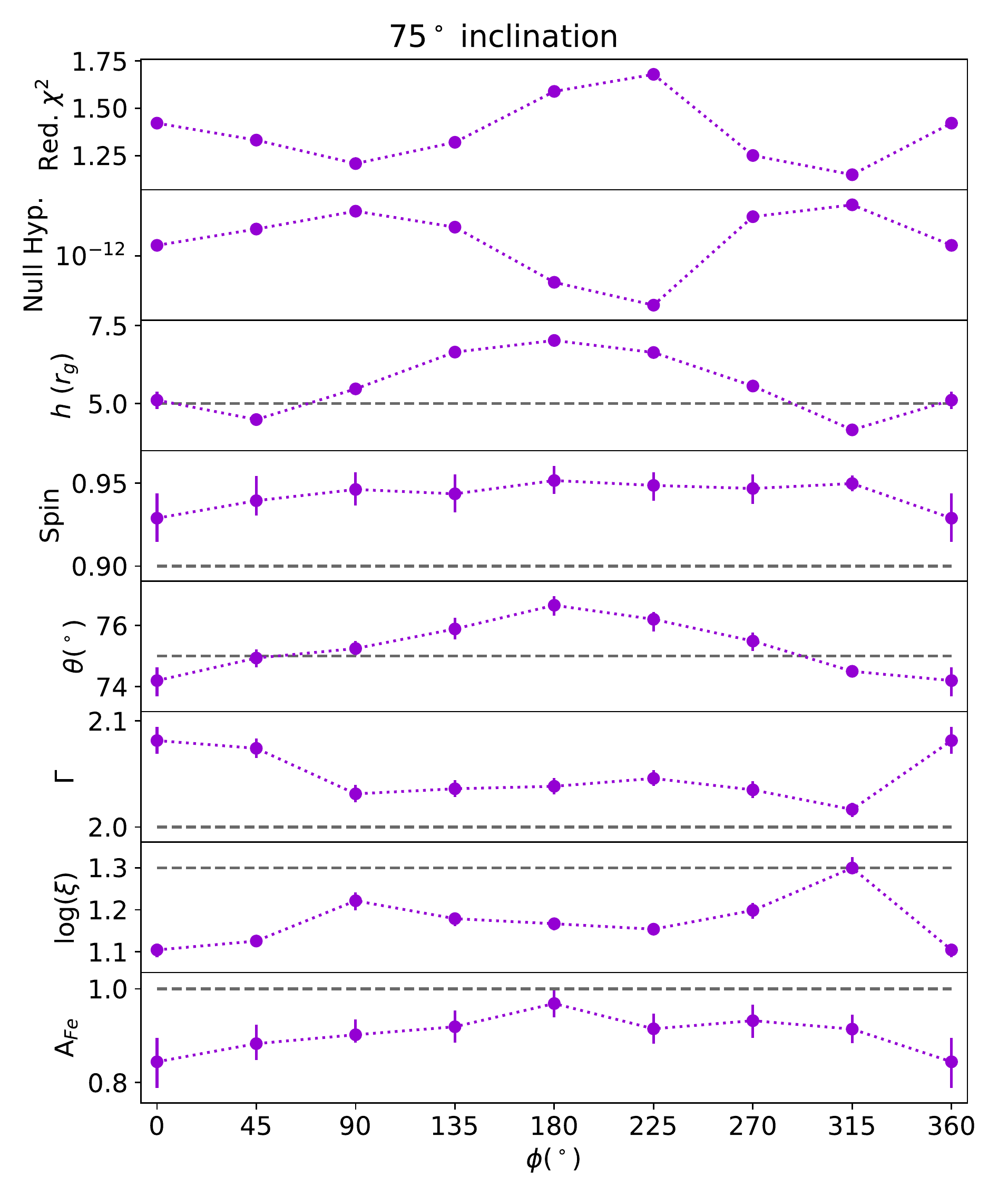}
    \caption{For a warped disk with tilt \eq{$\beta$}{\ang{15}}, we show the \xspec{} results obtained with \relxilllp{} for all $\phi$ values. In purple are the fit values with \SI{90}{\percent} confidence error bars; the points are connected by a dotted line for clarity. The dashed grey lines show the simulated parameter values. In the upper right plot, the red dashed curve shows the outer disk inclination. Each plot collects the results for one observer inclination. For \eq{$\phi$}{\ang{180}} at \ang{45} inclination (the middle point in the top right plot), \xspec{} was unable to calculate errors with all of these parameters varying freely. We instead froze the lamppost height at \SI{2}{\rg}. \explain{Added inclination titles to the four plots}}
    \label{fig:fitsT15}
\end{figure*}

We next fit our default warped disk model with \xspec{}. 
The plots of the results from these fits, including \SI{90}{\percent} confidence error bars, are in Figure \ref{fig:fitsT15}.

The goodness of the fit depends on $\phi$, with the best fit occurring around \eq{$\phi$}{\ang{90}} and/or \ang{270}, where the outer disk inclination is closest to the inner disk inclination.
For these azimuths at \eq{$\theta$}{\ang{25}}, we find that \h{}, $\Gamma$, $\log(\xi)$, and $A_{Fe}$ are all fit close to their true values.
The spin, is consistently fit too low, though it is still fit with high spin (\SIrange[range-phrase={ -- }]{\sim0.75}{0.85}{}).
This underestimation is greater than we found for the aligned disk, which was only lower than the true value by \SI{\sim0.05}{}.
At azimuths near \eq{$\phi$}{\ang{0}}, where the outer disk is more highly inclined than the inner disk, the reduced $\chi^2$ tends to be reasonably good, but the fit values poorly match the simulated values; this is particularly true at \eq{$\theta$}{\ang{25}}, where the height is half its true value and inclination is off by \ang{15}.
Interestingly, at the lowest observer inclination the fit inclination seems to track the outer disk inclination; this is shown by the red dashed curve in the \eq{$\theta$}{\ang{25}} plot.
All three other inclinations, however, are fit with values quite close to the inclination of the inner disk.

In the case of the \ang{45} inclination results in Figure \ref{fig:fitsT15}, we observe that all parameters except the lamppost height are fit rather well for the azimuths with good fits (\SIrange[range-phrase={ -- }]{\sim270}{90}{\degree}).
For \ang{60} inclination the best fit appears to be around \eq{$\phi$}{\ang{270}}.
Here, once again, the inclination, power law index, disk ionization, and iron abundance are fit very close to their true values, while the lamppost height and spin parameter are slightly off.
For \ang{75}, the \eq{$\phi$}{\ang{270}} is among the azimuths with the lowest $\chi^2$, though the best fit is actually at \ang{315}.
As predicted from the spectra in Figure \ref{fig:profile_r}, the spin is consistently overestimated because the profile has been broadened by the contribution of the outer disk.

The overall result seems to be that there is difficulty finding a good fit for out data with \relxilllp{} when the angular momenta axes of the two disks appear aligned to the observer (at \eq{$\phi$}{\ang{0}} and \eq{$\phi$}{\ang{180}}), likely because these viewing angles are where the tilt manifests to the observer entirely as difference in inclination.
For other azimuths, where the difference between the inclinations of the disks is smaller (i.e. close to \eq{$\phi$}{\ang{90} or \ang{270}}), there can be a reasonable fit with parameters close to their true value, especially at higher inclinations.

Next, we will explore the possibility of fitting our data with a two-component \relxilllp{} disk and see if it yields better results.

\subsection{Fitting with a Two-Component Disk}

\begin{figure}[t]
  \centering
  \includegraphics[width=\linewidth]{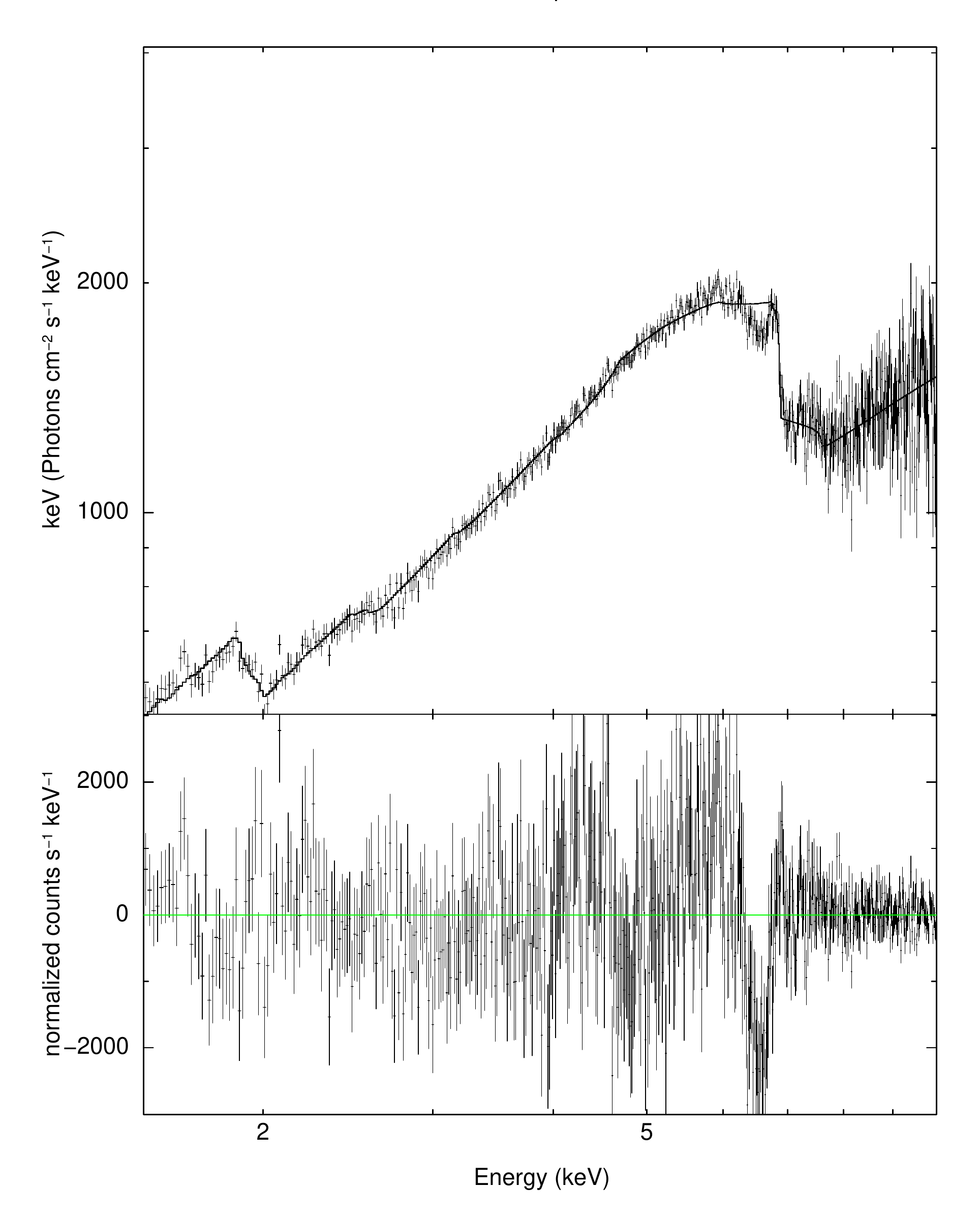}
  \caption{\xspec{} fit with a single \relxilllp{} disk model with \ang{25} inclination and \eq{$\phi$}{\ang{45}}.}
  \label{fig:singlefit}
\end{figure}

An obvious choice to improve the fitting of a warped accretion disk is to use a two-component \relxill{} model.
For an illustrative example, let us examine the \xspec{} fit for \ang{25} inclination from \ang{45} azimuth, shown in Figure \ref{fig:singlefit}.
The data shows a spur at \SI{~6.4}{\kev}, which a single \relxilllp{} model is unable to account for.
From this fit, we get the values displayed in Table \ref{tab:singlefit}.
In particular, the fit values for inclination and lamppost height are significantly off; worth noting, though, is that the inclination is fit with a value quite close to the inclination of the outer disk.

\begin{table}
  \renewcommand{\arraystretch}{1.5}
  \centering
  \begin{tabular*}{\linewidth}{l @{\extracolsep{\fill}} c c}
    \hline
    Parameter & Simulated Value & Fit Value \\
    \hline
    Reduced $\chi^2$ & --- & 1.414 \\
    Null hypothesis & --- & \num{2.13e-10} \\
    Lamppost height (\SI{}{\rg}) & $5$ & \err{ 2.006 }{ 0.0 }{ 0.145 }\\
    Spin & $0.9$ & $0.811 \pm 0.004$ \\
    Inclination(\SI{}{\degree}) & $25$ & \err{ 36.694 }{ 0.227 }{ 0.241 }\\
    PL Index & $2.0$ & $1.961 \pm 0.004$ \\
    log($\xi$) & $1.3$ & $1.320 \pm 0.002$ \\
    A$_{Fe}$ & $1.0$ & $1.300 \pm 0.043$ \\
    \hline
  \end{tabular*}
  \caption{Fit results using a single \relxilllp{} model for the reflection spectrum from a warped disk viewed by an observer at a \ang{25} inclination and \ang{45} azimuth.}
  \label{tab:singlefit}
\end{table}

We find that a two-component \relxilllp{} gives higher-accuracy estimates of the simulated system parameters than a one-component fit.
We perform the two-component fit by adding two \relxilllp{} models in \xspec{}.
We tie the spin, lamppost height, and power law index for the two \relxilllp{} models together.
For inclination, we allow both the inner disk and outer disk to be fit independently of one another.
Furthermore, we tie the outer radius of the first \relxilllp{} model to the inner radius of the second model, which gives us a measure of \rbp{}.
We do not fix the outer disk ionization and iron abundance to the inner disk values; we expect that the asymmetric illumination due to the lamppost being offset from the outer disk angular momentum and shadowing by the inner disk may cause these values to be poorly fit.
Thus, the values that we are fitting for the two component model are: lamppost height, BH spin, $\theta_{In}$, $\theta_{Out}$, \rbp{}, $\Gamma$, inner log($\xi$), inner A$_{Fe}$, outer log($\xi$), and outer A$_{Fe}$.

\begin{figure}[t]
    \centering
    \includegraphics[width=\linewidth]{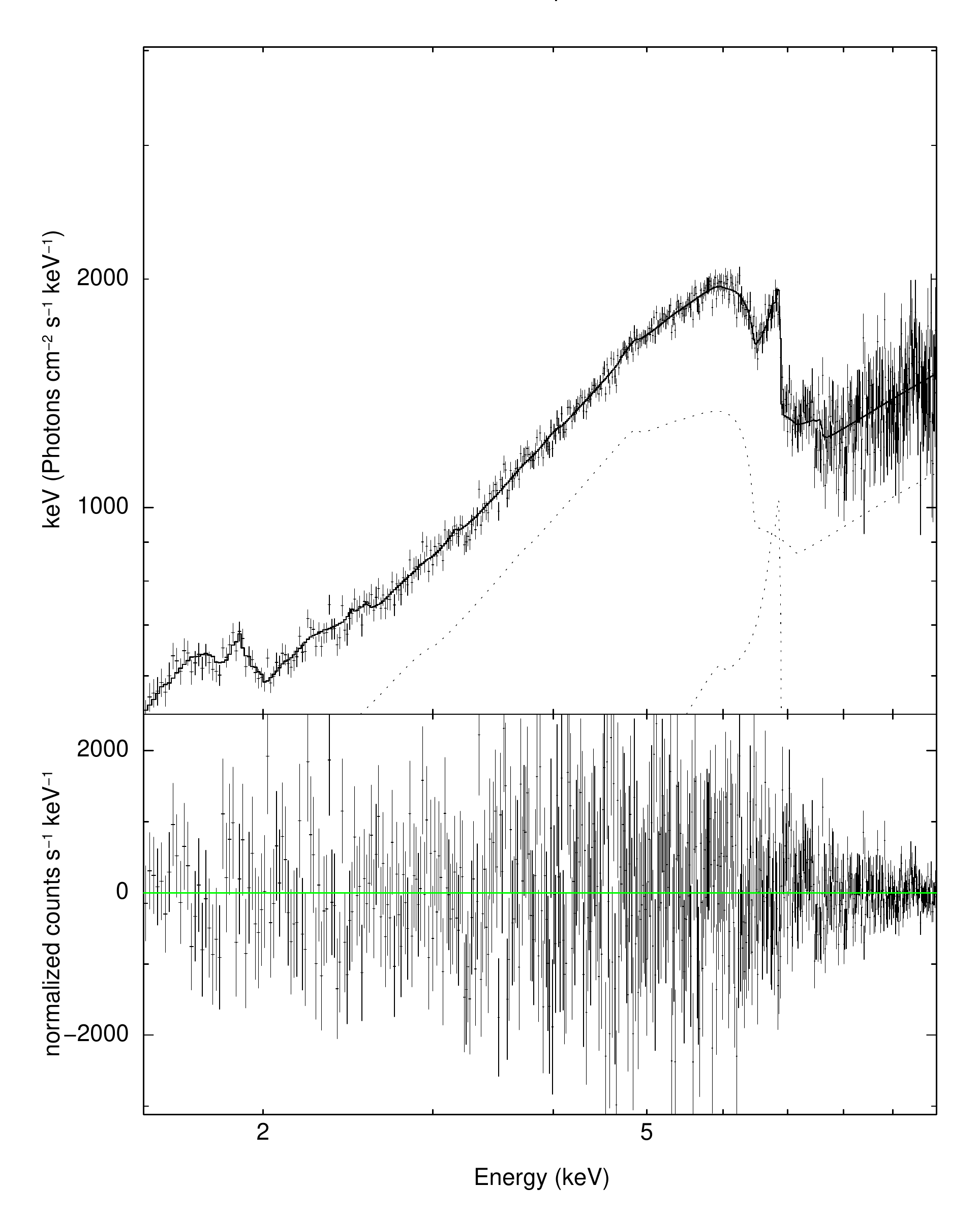}
    \caption{\xspec{} fit with a double \relxilllp{} disk model with \ang{25} inclination and \eq{$\phi$}{\ang{45}}.}
    \label{fig:doublefit}
\end{figure}

Figure \ref{fig:doublefit} and Table \ref{tab:doublefit} show the result of the two-component fit, which has much better residuals.
The lamppost height, spin, and inclination results are much closer to the simulated results than in the case of the one-component fit. 
Also noteworthy is that the fit recovers the location of the transition radius \rbp{} reasonably well, though not exactly.
In terms of disk composition, the iron abundance was slightly overestimated by the single \relxilllp{} model but is accurate for both the inner and outer disks in the double \relxilllp{} model.
The ionization, though, was accurate in the single \relxilllp{} model, but the double model accurately fits the outer ionization while slightly underestimating the ionization of both disks.

\begin{table}[h]
    \renewcommand{\arraystretch}{1.45}
    \centering
    \begin{tabular*}{\linewidth}{l @{\extracolsep{\fill}} c c}
        \hline
        Parameter & Simulated Value & Fit Value \\
        \hline
        Reduced $\chi^2$ & --- & 1.142 \\
        Null hypothesis & --- & \num{1.054e-1} \\
        Lamppost height (\SI{}{\rg}) & $5$ & \err{ 5.503 }{ 0.461 }{ 0.391 } \\
        Spin & $0.9$ & \err{ 0.881 }{ 0.027 }{ 0.027 }\\
        Inner disk inclination (\SI{}{\degree}) & $25$ & \err{ 26.476 }{ 0.535 }{ 0.547 }\\
        Outer disk inclination (\SI{}{\degree}) & $37.05$ & \err{ 36.491 }{ 0.563 }{ 0.501 }\\
        PL Index & $2.0$ &  \err{ 2.047 }{ 0.022 }{ 0.022 }\\
        Inner log($\xi$) & $1.3$ &  \err{ 1.144 }{ 0.082 }{ 0.159 }\\
        Inner A$_{Fe}$ & $1.0$ & \err{ 0.981 }{ 0.085 }{ 0.369 } \\
        \rbp{} (\SI{}{\rg}) & $15$ & \err{ 11.515 }{ 0.442 }{ 0.650 }\\
        Outer log($\xi$) & $1.3$ &  \err{ 1.170 }{ 0.117 }{ 0.146 }\\
        Outer A$_{Fe}$ & $1.0$ &  \err{ 1.002 }{ 0.224 }{ 0.944 }\\
        \hline
    \end{tabular*}
    \caption{Fit results using a double \relxilllp{} model for the reflection spectrum from a warped disk viewed by an observer at a \ang{25} inclination and \ang{45} azimuth.}
    \label{tab:doublefit}
\end{table}

\begin{figure*}
    \centering
    \includegraphics[width=0.49\linewidth]{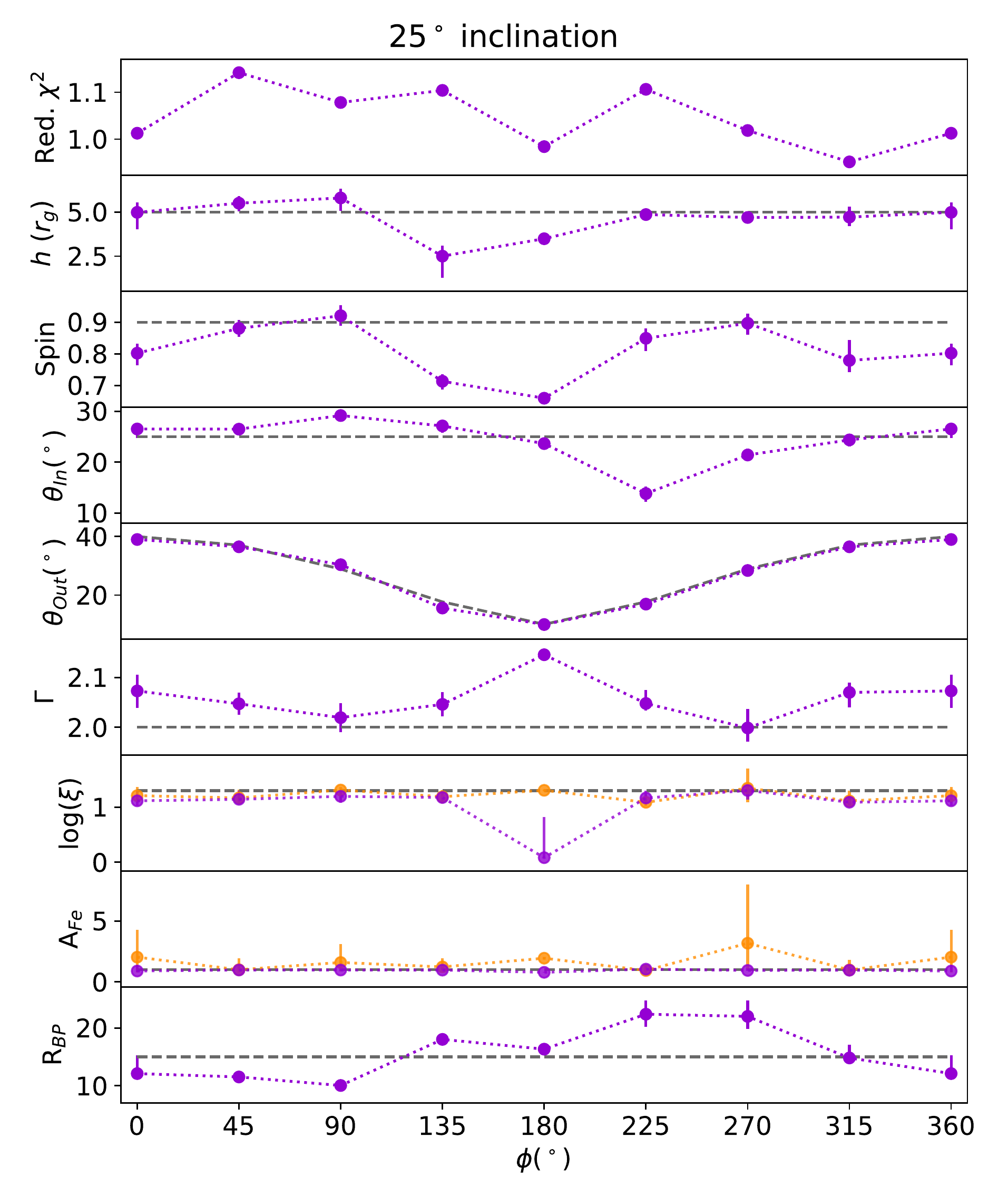}
    \includegraphics[width=0.49\linewidth]{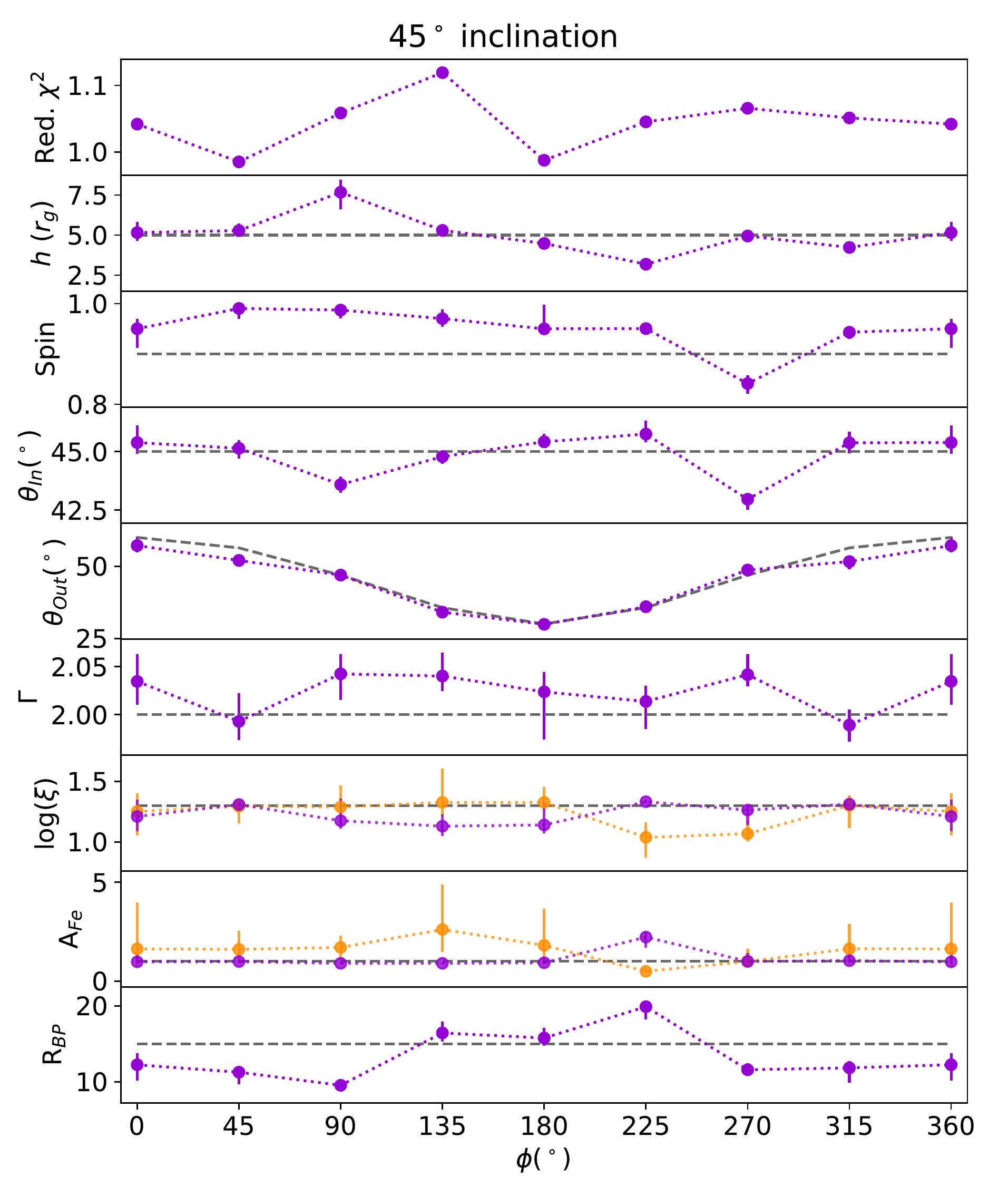}\\
    \includegraphics[width=0.49\linewidth]{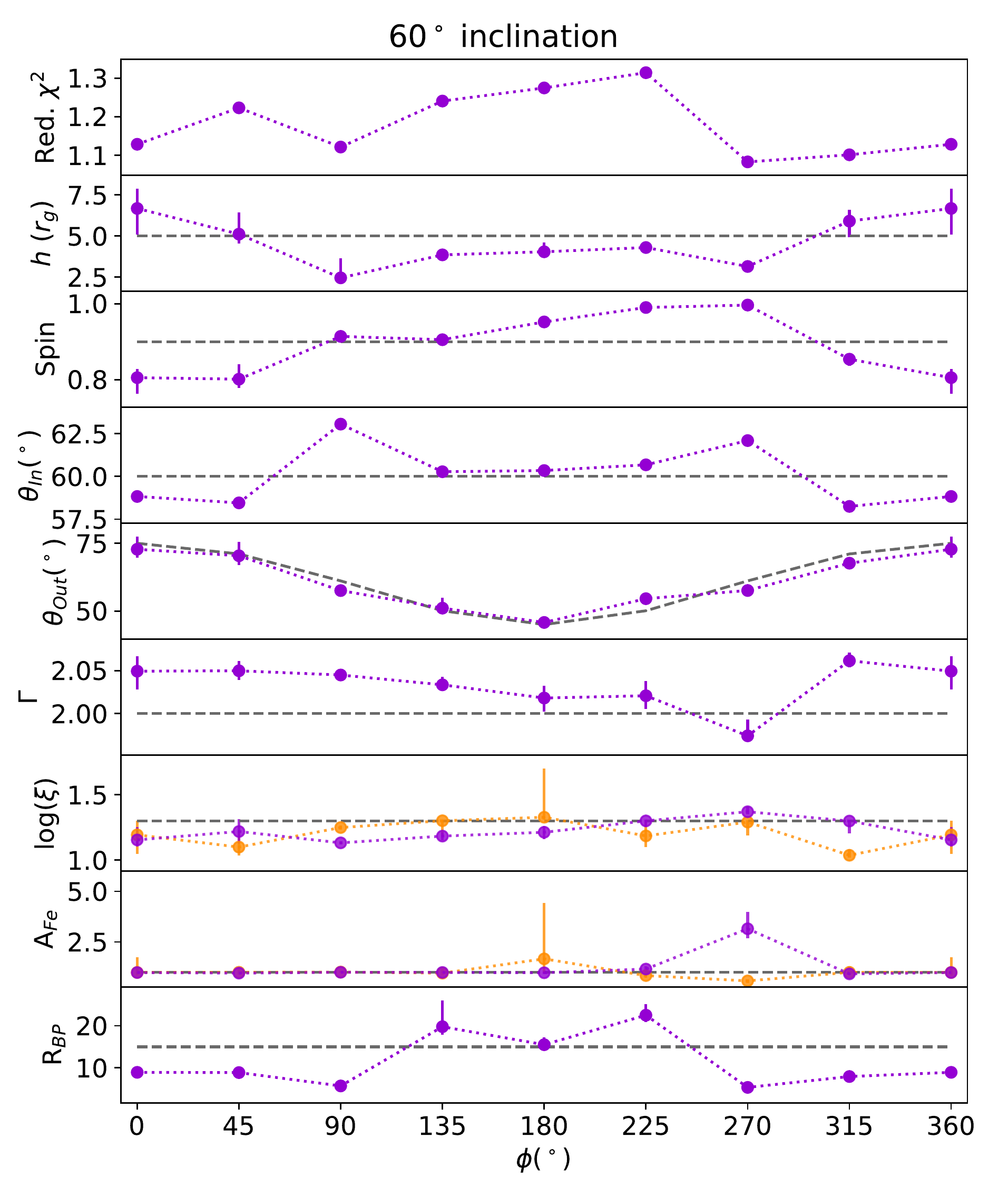}
    \includegraphics[width=0.49\linewidth]{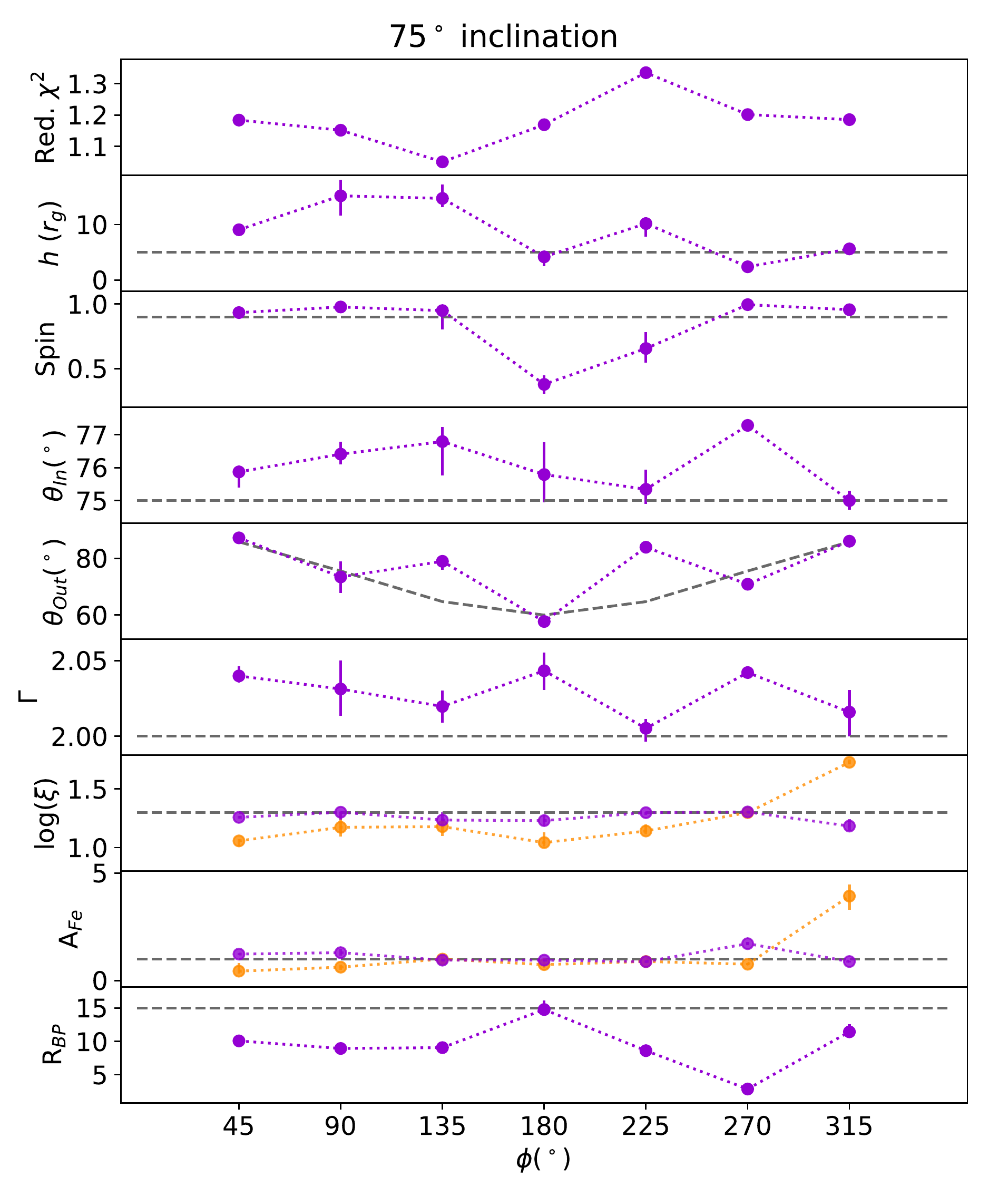}
    \caption{Results of fitting our simulated data with a double \relxilllp{} model in \xspec{}. For log($\xi$) and A$_{Fe}$, the purple points indicates the inner disk result and the orange indicates the outer disk. We do not fit \eq{$\phi$}{\ang{0}} at \ang{75} inclination (the bottom right plot) because the outer disk is viewed exactly edge on at \ang{90}, and the \relxilllp{} model is only valid up to \ang{87}. Null hypothesis values are omitted to make room for the additional fit results of outer disk inclination and \rbp{}. \explain{Added inclination titles to the four plots}}
    \label{fig:fitsT15double}
\end{figure*}

We have fit the simulated data for all azimuths with the double \relxill{} model as in Figure \ref{fig:doublefit}, and plot the results in Figure \ref{fig:fitsT15double}.
In most cases, the two-component model fits the simulated data much better than the one-component model, specifically with regards to the inclination.

\section{Conclusions}
\label{sec:conclusion}

In this paper, we have shown the effect that disk warping has on the profile of the fluorescent iron line from BH accretion disks.
Though we have focused on AGN, our method is valid for binary black holes as well.

We have estimated the impact of multiple scatterings, finding that their contribution to the iron line flux is stronger for warped disks than unwarped disks.
The impact of the multiply scattered photons on the iron line shapes seems to be small though. 
Utilizing the new \xillverrr{} results \citep{wilkins2020}, which calculate the reflection spectra of radiation returning to the disk (i.e. two scatterings) will allow for a more accurate treatment of the reflection spectra of warped accretion disks.
More detailed studies taking into account the impact of the radiation on the ionization state of the disk photosphere are needed to quantify the effect of multiply scattered photons on the inferred spin and inclination parameters.

We find that fitting the warped disk data with a standard single disk \relxill{} model can lead to under or overestimating the black hole spin and inclination parameters by tens of percents, especially for viewing angles near \eq{$\phi$}{\ang{0}} and \eq{$\phi$}{\ang{180}} where the warp leads to very different inclinations of the two disks.
In our fits of a warped disk, the spin parameter was off by as much as \num{0.2} in some cases.
We expect that for smaller values of \rbp{} and larger values of $\beta$ that this will be off by even more.
We showed in Figures \ref{fig:profile_r} and \ref{fig:profile_beta} the effect that varying warp radius and warp angle has on the iron line profile from a warped disk.
For warp radii all the way out to \SI{50}{\rg} we see a visible change in the profile.
For warp angle, the effect becomes noticeable at even \ang{5} and significant by \ang{15}. 
For systems at low inclinations, the fit value of inclination tends to match the outer disk, while at higher inclinations it fits to the inner disk inclination.

We have shown that by using a two-component \relxilllp{} model, the spin, inclination, and corona height can be estimated with higher accuracies, and that this method can also estimate the inclination of the outer disk and the radius of the warp between the inner and outer disks.

Being able to fit spectra from warped accretion disks will become more important for the analysis of the high-throughput, excellent energy resolution energy spectra that the upcoming {\it XRISM} and {\it Athena} missions will deliver. 
Even though our simulations were done with the resolution of {\it XRISM} and {\it Athena}, the disk warping may already show in the data from current satellites.
This is particularly true for systems for which the inclinations of the two disk components towards the observer differ significantly (e.g. \ang{0} at \ang{25} inclination in Figure \ref{fig:profile_az}). 

There are several potential targets for observation which may exhibit reflected power law emission from warped accretion disks.

Using data from NICER, \cite{miller2018} found that MAXI J1535-571 is best fit with a combination of \relxill{} and \texttt{relline}, the relativistic blurring portion of \relxill{}.
The \texttt{relline} source is far from the black hole (it has a radius of \err{144}{60}{140}\SI{}{\rg}) and is not well fit by either \relxill{} or \xillver{}.

\cite{krawcz2019} investigated the gravitationally lensed quasar RX J1131-1231, and were unable to fully explain the observed line energies with microlensing by the lensing galaxy; a warped disk offers a possible alternate explanation for the observed lines.

\cite{ingram2017} explain the quasi-periodic oscillation (QPO) from H~1743-322 as the result of asymmetric illumination of the disk, which may be a result of disk warping. 
Their best fit result gives an inner disk radius of $31.47^{+5.83}_{-3.66}$ \SI{}{\rg}.
This could be the location of \rbp{}, or it could be that the inner flow within this radius is not dense enough to contribute significantly to the iron fluorescence. 
Based on their form of Equation \ref{eq:iout}, they show that \eq{$\beta$}{\ang{15}} is consistent with the data within $0.5\sigma$.

Recently, \cite{connors2019} analysed ASCA+RXTE data of XTE J1550-564 during a 1998 outburst in the hard-intermediate state.
Using the latest \relxill{} models, they found the inclination of the reflection region of the disk to be \ang{40}, disagreeing with previous measurements that the jet is inclined at \ang{75} \citep{orosz2011}.
Though disk warping may be the cause of this disagreement, they suggest alternatives such as a thick inner disk.
Further work confirms this low inclination, suggesting that XTE J1550-564 may indeed contain a warped disk \citep{connors2020}.

Fitting data from RX J1131-1231, H~1743-322 and XTE~J1550-564 data with the warped disk models described in this paper would be a worthwhile activity.

\acknowledgments
The authors thank NASA for the support through the grants NNX16AC42G and 80NSSC18K0264, as well as the Washington University McDonnell Center for the Space Sciences. Additionally, they would like to thank Fabian Kislat for his work in speeding up the ray-tracing code used.
\hspace{2em}

\appendix

\section{Results from fitting an unwarped disk}
\label{ap:fits}
\begin{figure*}[b]
    \centering
    \includegraphics[width=0.47\linewidth]{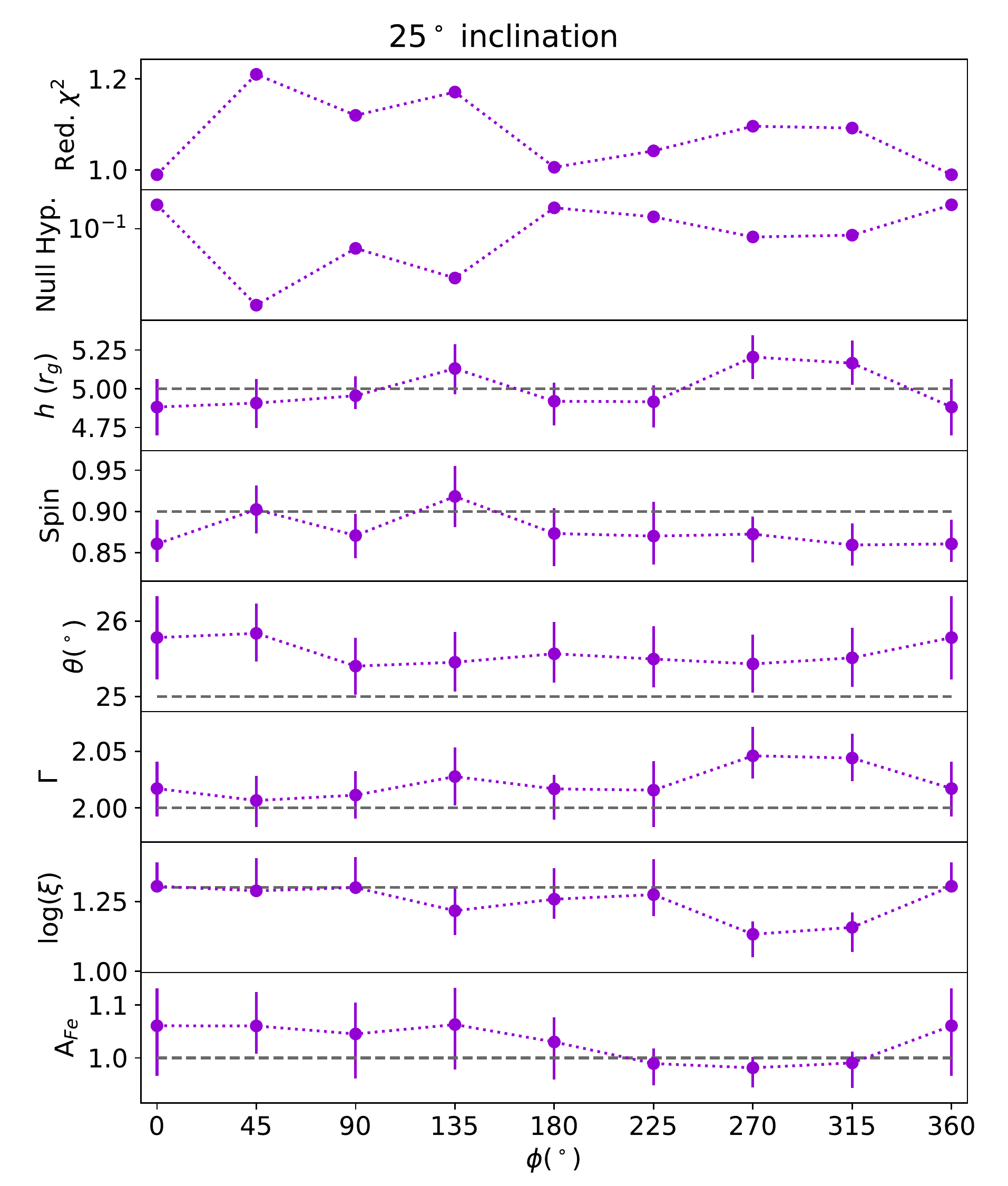}
    \includegraphics[width=0.47\linewidth]{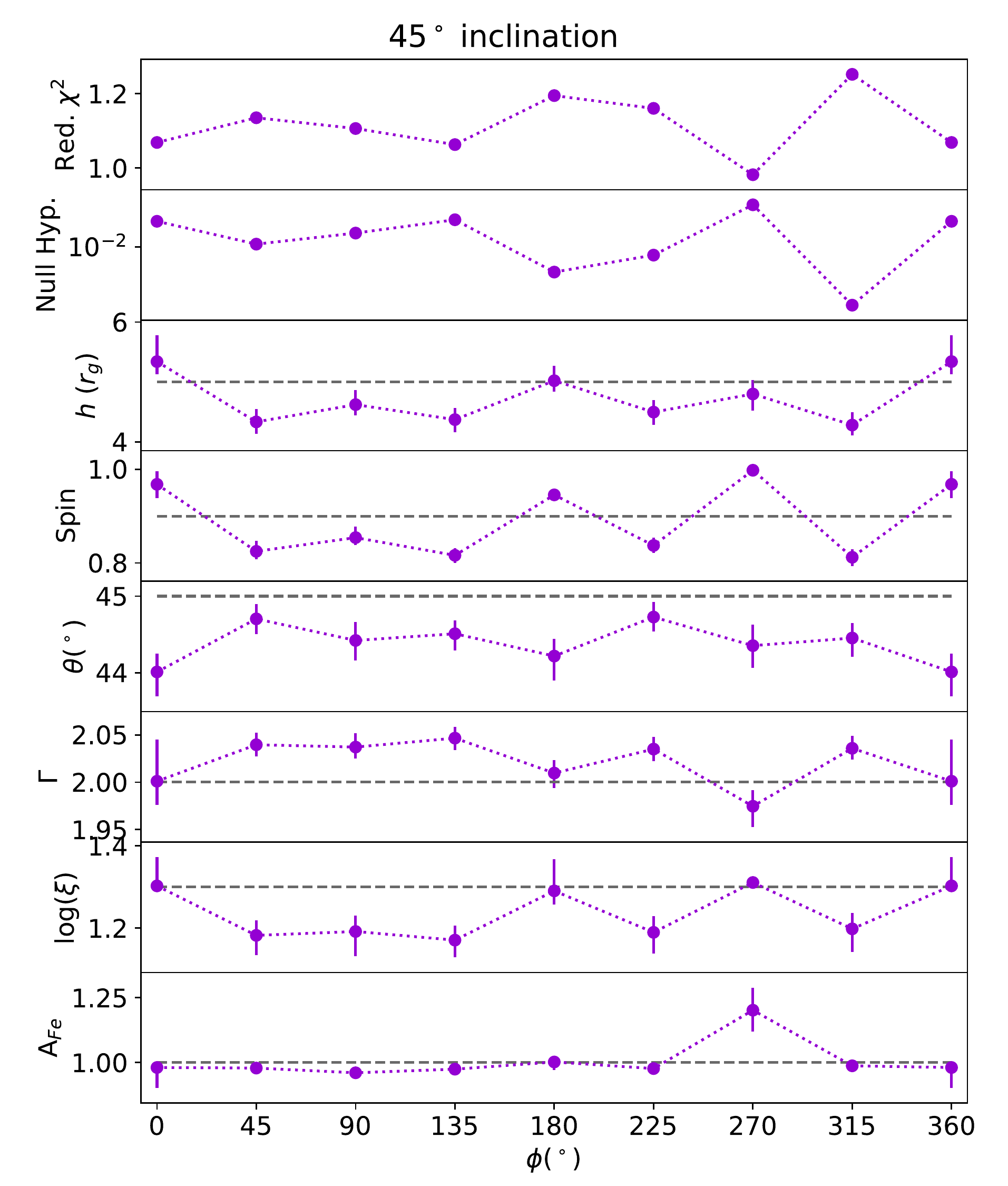}\\
    \includegraphics[width=0.47\linewidth]{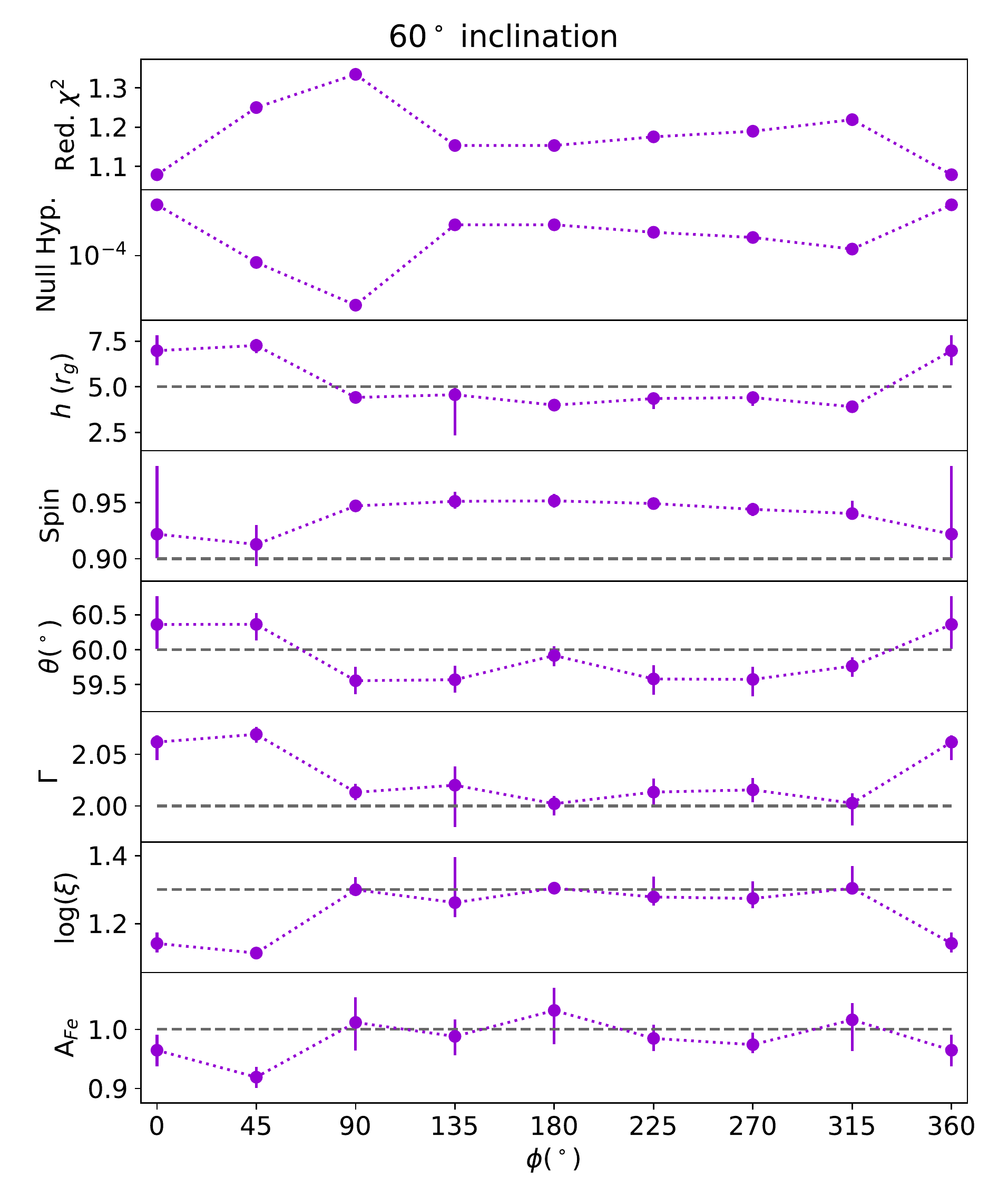}
    \includegraphics[width=0.47\linewidth]{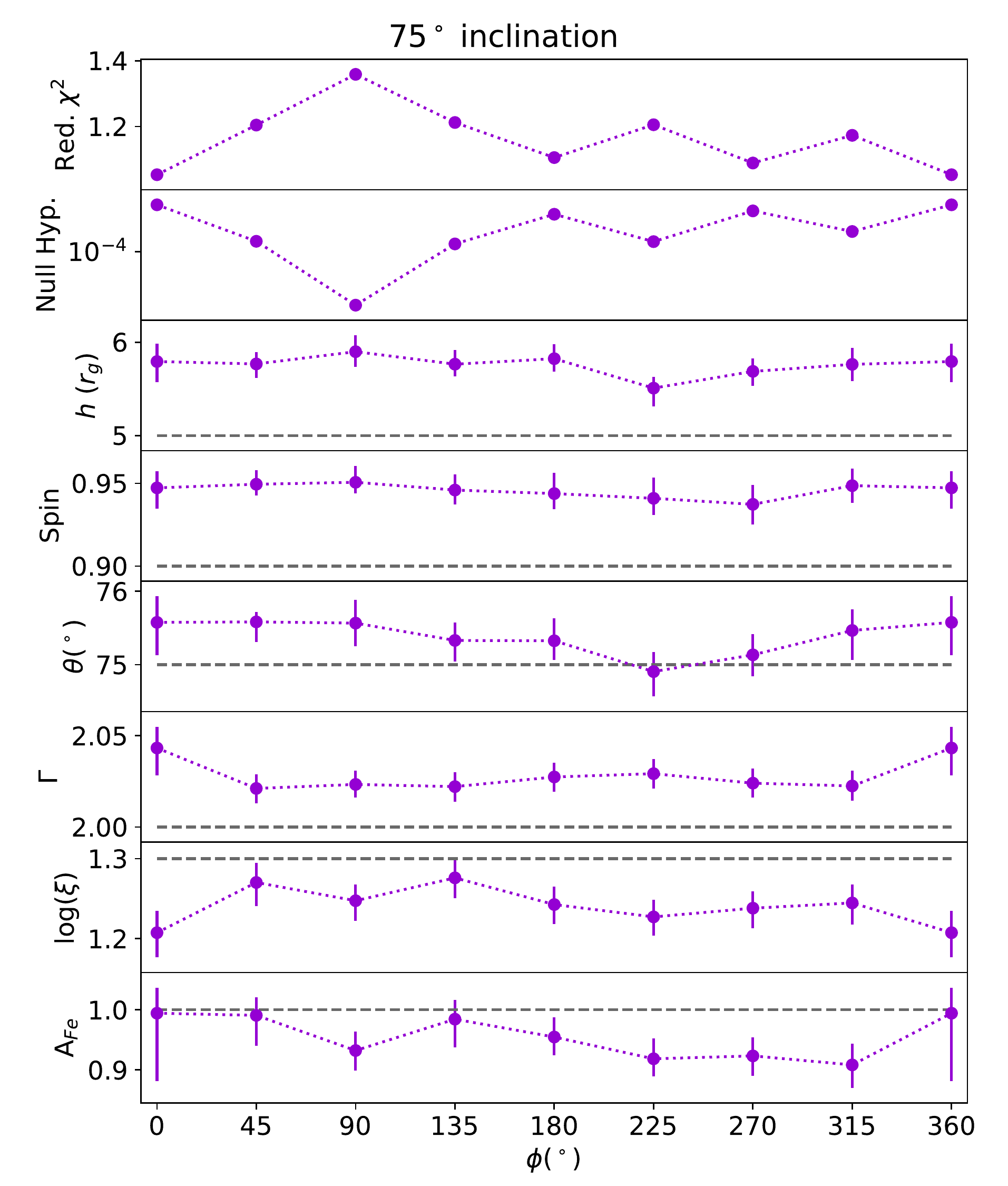}
    \caption{For an unwarped disk, the \relxilllp{} fit results for all observers. Formatting is the same as Figure \ref{fig:fitsT15}. \explain{Added inclination titles to the four plots}}
    \label{fig:fitsT0}
\end{figure*}

\end{document}